\documentclass[12pt, onecolumn]{IEEEtran}

\usepackage{cite}
\usepackage{pslatex,bm}
\usepackage{epsfig}
\usepackage{amsmath, amssymb, latexsym}
\usepackage{amsthm}
\usepackage{array}
\usepackage{multicol}
\usepackage{color}
\usepackage{url}
\usepackage{dsfont}
\usepackage{caption}
  \usepackage{comment}
  \usepackage{lipsum}
  \usepackage{amsfonts}
\usepackage{booktabs}
\usepackage{siunitx}
  \usepackage{geometry}
  \usepackage{cases}
  \usepackage{tikz}
  \usepackage[colorinlistoftodos,prependcaption,textsize=small]{todonotes}

\psfull
\newtheorem{lemma}{Lemma}
\newtheorem{rmk}{\sf Remark}
\newtheorem{definition}{Definition}

\newtheorem{Prop}{Proposition}

\newcommand{\QEDA}{\hfill\ensuremath{\square}}%
\newcommand{\ind}{\mathbin{\rotatebox[origin=c]{90}{$\vDash$}}}
\newcommand{\Var}{\mathop{\rm Var}}

\newcommand{\textsfP}{{\normalfont \textsf{P}}}
\newcommand{\textsfR}{{\normalfont \textsf{R}}}
\newcommand{\textsfC}{{\normalfont \textsf{C}}}
\newcommand{\textsfV}{{\normalfont \textsf{V}}}
\newcommand{\textsfM}{{\normalfont \textsf{M}}}

\newcommand{\textsfVG}{{\normalfont\textsf{V}_G}}
\newcommand{\textsfbarV}{{\normalfont\bar{\textsf{V}}}}
\newcommand{\textsfbarC}{{\normalfont\bar{\textsf{C}}}}
\newcommand{\SNR}{{\normalfont \textsf{SNR}}}

\DeclareMathAlphabet{\mathpzc}{OT1}{pzc}{m}{it}

\newcommand{\eps}{\epsilon}
\renewcommand{\baselinestretch}{2}

\def\b0{{\mathbf 0}}

\newtheorem{Theorem}{Theorem}

\begin{document}
\newcommand{\CLASSINPUTtoptextmargin}{1.72in}
\title{Gaussian Broadcast Channels under Heterogeneous Blocklength Constraints}

\author{\IEEEauthorblockN{Pin-Hsun Lin\dag, Shih-Chun Lin\ddag, Peng-Wei Chen\S, Marcel Mross\dag, and \\
\vspace{-0.4cm}Eduard A. Jorswieck\dag\\
\vspace{-0.4cm}\dag Institute for Communication Technology,\\
\vspace{-0.4cm} Technische Universit\"{a}t Braunschweig, Germany\\
\vspace{-0.4cm}\ddag Department of Electrical Engineering, National Taiwan University, Taiwan,\\
\vspace{-0.4cm}\S Department of Electrical and Computer Engineering,\\
 \vspace{-0.4cm}National Taiwan University of Science and Technology, Taiwan,\\
 \vspace{-0.4cm}Email:{\{Lin, Mross, Jorswieck\}@ifn.ing.tu-bs.de,\\
 \vspace{-0.4cm}  sclin2@ntu.edu.tw, M10802280@mail.ntust.edu.tw}}
\thanks{ Part of the work is presented in ISIT 2021 \cite{PHL_ISIT21} and ICC 2022 \cite{PHL_ICC22_ED}. }
 }

\maketitle


\vspace{-0.8cm}

\begin{abstract}
Future wireless access networks aim to simultaneously support a large number of devices with heterogeneous service requirements, including data rates, error rates, and latencies. While achievable rate and capacity results exist for Gaussian broadcast channels in the asymptotic blocklength regime, the characterization of second-order achievable rate regions for heterogeneous blocklength constraints is not available. Therefore, we investigate a two-user Gaussian broadcast channel (GBC) with heterogeneous blocklength constraints, specified according to users' channel output signal-to-noise ratios (SNRs). We assume the user with higher output SNR has a shorter blocklength constraint. We show that with sufficiently large output SNR, the stronger user can perform the \textit{early decoding} (ED) technique to decode and subtract the interference via successive interference cancellation (SIC). To achieve it, we derive an explicit lower bound on the necessary number of received symbols for a successful ED, using an independent and identically distributed Gaussian input. A second-order rate of the weaker user who suffers from an SNR change due to the heterogeneous blocklength constraint, is also derived. Numerical results show that ED can outperform the hybrid non-orthogonal multiple access scheme when the stronger channel is sufficiently better than the weaker one. Under the considered setting, about 7-dB SNR gain can be achieved. These results shows that ED with SIC is a promising technique for the future wireless networks.
\end{abstract}


\section{Introduction}\label{Sec_Intro}
Ultra-reliable and low-latency communication (URLLC) is one of the target application scenarios in 5G and beyond \cite{3GPP_URLLC,3GPP_URLLC2,Durisi_URLLC}, which has attracted many research efforts. One important branch of research is finite blocklength analysis \cite{Polyanskiy_finite_block_length}, which has been extended to different multiuser cases such as the multiple access channel \cite{Ebra_IT15} and \cite{Scarlett_GMAC_DMS}, asymmetric broadcast channel \cite{Tan_MUC_dispersion}, Gaussian broadcast channel \cite{Gorce_GBC_dispersion} (GBC), the strong converse for GBC \cite{Tan_strong_converse_GBC}, GBC with hard deadline \cite{Sheldon_GBC_FBL}, and the channel with state \cite{Scarlett_GPC}. For 5G, the flexibility for resource allocation was obtained by fine numerology \cite{Korrai_5G}. Within each service class, most studies considered homogeneous blocklength and latency constraints among the users. In practice, heterogeneous blocklengths among users should be considered due to different requirements of latency, quality of service, and different channel conditions among users. Therefore, we investigate heterogeneous blocklengths among users.

The GBC with heterogeneous blocklength constraints has been studied in \cite{xu_hybrid_NOMA}, \cite{Tuninetti_GBC_hard_deadline}. In \cite{xu_hybrid_NOMA}, hybrid non-orthogonal multiple access (HNOMA) is analyzed, where treating interference as noise (TIN), normal superposition coding (homogeneous blocklength at all users) with successive interference cancellation (SIC), and time-division multiple access (TDMA) are used in this case. In particular, the weaker user's codeword is divided into two shorter codewords, where one of them has the same blocklength as that of the stronger user. Therefore, the normal superposition coding with SIC can be applied.  In \cite{Tuninetti_GBC_hard_deadline}, after TDMA transmission, a user is forced to decode both the intended and interference codewords. Such decoding is similar to decoding a common message in a GBC, and the rate is limited by the link with the lower output SNR. Aside from the above, one unsolved but important issue is how to perform superposition coding with SIC when the two users use codewords of different lengths. More specifically, the performance of SIC is unclear for the user with a higher output signal-to-noise ratio (SNR) but shorter blocklength in a two-user GBC.

 In contrast to \cite{xu_hybrid_NOMA}, \cite{Tuninetti_GBC_hard_deadline}, we argue that the stronger user with sufficiently large output SNR can still decode the interference to perform SIC based on the partially received symbols. The key is the early decoding (ED) technique: assume that a code is designed for a channel with a specific output SNR under a specific error probability and blocklength. Via ED, the message can be decoded using only a certain fraction of the codeword when it is transmitted through a channel with a larger output SNR. This concept has been investigated with traditional first-order asymptotic analysis in \cite{Azarian_ED}, which tells us that in a GBC, a user with higher output SNR than the other can decode successfully using fewer received symbols. The concept of ED has already been used in several wireless scenarios, not only to improve the latency performance, but also to increase the throughput of a network. A popular application of ED is in cognitive radio (CR) \cite{Jovicic_CR}. In addition to \cite{Azarian_ED}, there are further works about ED: In \cite{Hou_EarlyDecoding}, ED is applied to short message noisy network coding under the same asymptotic assumption as in \cite{Azarian_ED}. In \cite{Sahin_ED}, the authors consider the necessary number of symbols for ED for binary input channels via numerical simulations under a finite blocklength assumption. Note that the analytical result for the finite blocklength regime of ED is missing in the above references, which motivates our work in this paper.

Our main contributions are as follows:
\begin{itemize}
  \item We consider a two-user GBC with heterogeneous blocklength constraints and two private messages (no common message), while the SIC is applied at the user having higher output SNR with a shorter (stricter) blocklength constraint,
  \item We derive a second-order lower bound on the necessary number of received symbols such that ED works as the first step of SIC, i.e., decoding the interference, while fulfilling the input power and error probability constraints. In particular, we analyze the dependence testing (DT) bound \cite{Polyanskiy_finite_block_length} over a fixed code instead of using Shannon's random coding scheme, to ensure that a specific code can be decoded when being transmitted through two different channels while fulfilling the aforementioned constraints. By applying ED, the stronger user can avoid using TIN as \cite{xu_hybrid_NOMA} and the stronger user's rate can be improved. We also derive the second-order rate of the weaker user, whose received symbols encounter an SNR change due to the heterogeneous blocklengths.
  \item Based on the derived second-order rates of the stronger and weaker users in a GBC, we formulate the rate region problems for ED and HNOMA. In particular, we start the investigation from  individual power constraint (IPC) and generalize it to sum power constraint (SPC), and solve the corresponding programming problems numerically.
  \item Numerical results show that ED can significantly reduce latency in the finite blocklength regime. Under the considered setting, more than 10dB SNR gain can be achieved. In addition, ED partly outperforms HNOMA regarding the sum rate and the rate region, when the stronger user has a sufficiently better channel than the weaker user. Therefore, a hybrid system combining ED and HNOMA is the best-known achievable scheme.
\end{itemize}

This paper is organized as follows. Section \ref{Sec_intro} introduces the system model and preliminaries. Section \ref{Sec_MainResult} shows our main result: the minimum number of received symbols for a successful ED. In addition, we derive the weaker user's second-order rate, under both IPC and SPC. In Section \ref{Sec_simulation} we show the performance improvements numerically. We conclude this paper in Section \ref{Sec_conclusion}.\\

{
\renewcommand{\baselinestretch}{1.5}
\emph{Notation}$\colon$ Upper/lower case normal letters denote random/deterministic variables. Upper case calligraphic letters denote sets. The notation $a_i^j$ denotes a row vector $[a_i,\,a_{i+1},\,\ldots,\,a_{j}]$ while $a_1^j$ is simplified to $a^j$. We denote the inner product of two vectors $a^j$ and $b^j$ by $\langle a^j,b^j \rangle$. The probability of event $\mathcal{A}$ is denoted by Pr$(\mathcal{A})$. The expectation and variance are denoted by $\mathds{E}[\cdot]$ and $\mbox{Var}[\cdot]$, respectively. We denote the probability density function (PDF) and cumulative distribution function
(CDF) of a random variable $X$ by $f_X$ and $F_X$, respectively. The random variable $X$ following the
distribution with CDF $F$ is denoted by $\,X\sim \,F$. $\mathrm{Unif}(a,b)$ denotes the uniform distribution between $a\in\mathds{R}$ and $b\in\mathds{R}$. We use $X\ind Y$ to denote that $X$ and $Y$ are stochastically independent. The logarithms used in the paper are all with respect to base 2. We define $\textsfC(x)\triangleq\frac{1}{2}\log(1+x)$. Real additive white Gaussian noise (AWGN) with zero mean and variance $\sigma^2$ is denoted by $\mathcal{N}(0,\sigma^2)$. We denote the indicator function and identity matrix with dimension $n$ by $\mathds{1}$ and $\bm I_{n}$, respectively. We denote the inverse $Q$-function by $Q^{-1}(.)$ and the big-O and small-o by $\mathcal{O}(.)$ and ${o}(.)$, respectively.
}


\section{System Model and Preliminaries}\label{Sec_intro}
\subsection{System Model}
We consider a two-user GBC, where only private messages, but no common message, are transmitted to each user. Denote the blocklength of user $k$ by $n_k\in\mathds{N}^+$, $k=1,\,2$. We assume $n_1\geq n_2$. The received signal at user $k$ at time $i$ is expressed as follows:
\begin{align}
Y_{k,i}&=
\sqrt{h_k}X_{i}+Z_{k,i},\,i\in\{1,\ldots,n_k\},\,k=1,\,2,\,\label{EQ_BC_1st_ch_1}
\end{align}
where the channel input $X_i\in \mathcal{F}_{\ell}^{n_1}\subseteq\mathds{R}^{n_1}$, $\ell=\{IPC,\,SPC\}$, while $\mathcal{F}_{SPC}^{n_1}$ and $\mathcal{F}_{IPC}^{n_1}$ are sets of feasible codewords satisfying the upcoming power constraints in \eqref{EQ_SPConstraint} and \eqref{EQ_max_power_constraint_BC}, respectively, $Z_{1,i}\sim\mathcal{N}(0,1)$ and $Z_{2,i}\sim\mathcal{N}(0,1)$, are independent and identically distributed (i.i.d.) and mutually independent additive white Gaussian noises. Define two message sets $\mathcal{M}_k:=\{1,\ldots,\textsf{M}_k\},\,k=1,\,2$ for each user. Assume that the message tuple $(m_1,m_2)$ is uniformly selected from $\mathcal{M}_1\times\mathcal{M}_2$. The considered code $(\textsf{M}_1,\,\textsf{M}_2,\,n_1,\,n_2,\, \epsilon,\,\mathcal{F}_{\ell}^{n_1})$ consists of
\begin{itemize}
  \item two message sets $\mathcal{M}_k=\{1,\ldots,\textsf{M}_k\},\,k=1,\,2$,
  \item one encoder $f$: $\mathcal{M}_1\times \mathcal{M}_2\mapsto \mathcal{F}_{\ell}^{n_1},\,\ell\in\{IPC,\,SPC\}$,
  \item two decoders $\phi_k$: $\mathds{R}^{n_k}\mapsto \mathcal{M}_k$, $\,k=1,\,2$,
\end{itemize}
such that
\begin{align}\label{EQ_joint_Pe}
P_e^{n_1,n_2}:=\frac{1}{\textsf{M}_1\textsf{M}_2}\sum_{(m_1,m_2)=(1,1)}^{(\textsf{M}_1,\textsf{M}_2)}\mbox{Pr}(\hat{m}_1\neq m_1\mbox{ or }\hat{m}_2\neq m_2|(m_1,m_2)\mbox{ is sent})\leq \epsilon,
\end{align}
where $\epsilon\in(0,1)$ is a error probability constraint. Given $n_1$, $n_2$, and $\epsilon$, a message-size tuple $\left(\textsf{M}_1,\,\textsf{M}_2\right)$ is achievable, if an $(\textsf{M}_1,\,\textsf{M}_2,\,n_1,\,n_2,\,\epsilon,\,\mathcal{F}_{\ell}^{n_1})$-code satisfying \eqref{EQ_joint_Pe} exists. We consider the maximal power constraint on the channel input by defining the following set:
\begin{align}\label{EQ_SPConstraint}
\mathcal{F}_{SPC}^{n_1}:=\left\{x^{n_1}:\,||x^{n_1}||^2\leq n_1\textsf{P}\right\}.
\end{align}
An encoding error is declared if a generated codeword $x^{n_1}$ does not belong to $\mathcal{F}^{n_1}_{SPC}$.

\begin{figure}[h!]
\centering \epsfig{file=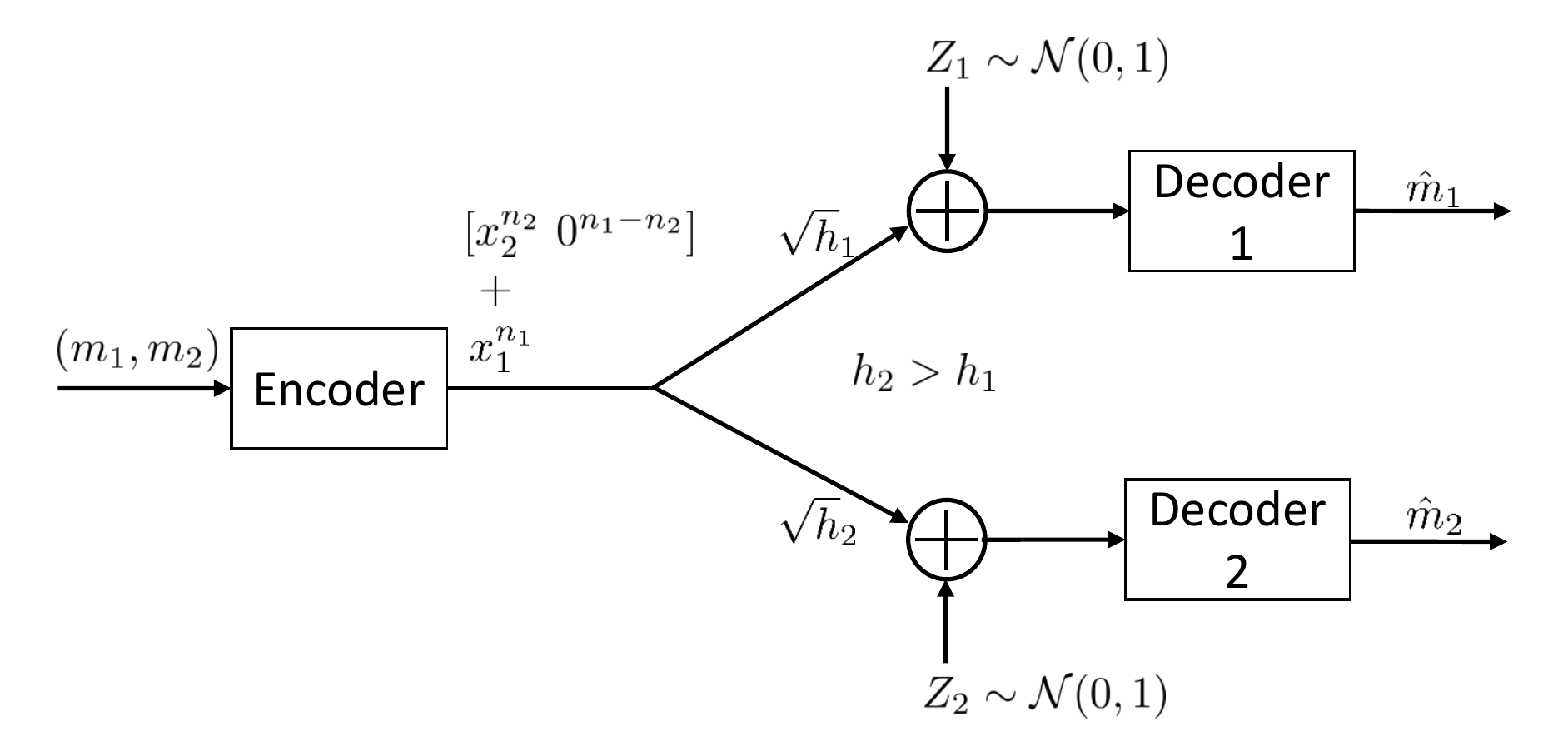, width=0.7\textwidth}
\caption{The considered 2-user GBC with heterogeneous blocklength constraints.}
\label{Fig_sys}
\end{figure}

To implement the encoder and decoder for the above model, we consider the superposition coding with successive interference cancellation (SIC), summarized in Fig. \ref{Fig_sys}. The received signals at the two receivers can be respectively expressed as follows:
\begin{align}
Y_{k,i}=\begin{cases}
&\hspace{-0.3cm}\sqrt{h_k}(X_{1,i}+X_{2,i})+Z_{k,i},\,\,\,i\in\{1,\ldots,n_2\},\label{EQ_BC_1st_ch_1}\\
&\hspace{-0.3cm}\sqrt{h_k}X_{1,i}+Z_{k,i},\,\hspace{1.5cm}i\in\{n_2+1,\ldots,n_1\},
\end{cases}
\end{align}
where the codewords of the two codes $\{X_{1}^{n_1}(m_1):\,m_1\in\mathcal{M}_1\}$ and $\{X_{2}^{n_2}(m_2):\,m_2\in\mathcal{M}_2\}$ are generated according to i.i.d. Gaussian distributions:  $X_{k,i}\sim\mathcal{N}(0,\textsf{P}_k),\,i\in\{1,\ldots,n_k\}$, $k=1,\,2$, and $\{X_{1}^{n_1}(m_1):\,m_1\in\mathcal{M}_1\}$ and $\{X_{2}^{n_2}(m_2):\,m_2\in\mathcal{M}_2\}$ are mutually independent.

In addition to \eqref{EQ_SPConstraint}, in the following analysis we also consider the maximal power constraint on each user's codewords, namely, the IPC, by defining the following set:
\begin{align}\label{EQ_max_power_constraint_BC}
\mathcal{F}^{n_1}_{IPC}:=\left\{x^{n_1}:\,x^{n_1}=x_1^{n_1}+[x_2^{n_2},\,{0}^{n_1-n_2}],\,||x_k^{n_k}||^2\leq n_k\textsf{P}_k\right\},\,\,k=1,\,2,
\end{align}
where ${\textsf{P}}_k$ is the power constraint at user $k$. An encoding error is declared if a generated codeword $x_1^{n_1}$ or $x_2^{n_2}$ does not belong to $\mathcal{F}^{n_1}_{IPC}$. In contrast to the IPC described by \eqref{EQ_max_power_constraint_BC}, we also call \eqref{EQ_SPConstraint} the SPC.

We consider the following decoding schemes. At the stronger user (user 2), the decoder first finds a unique $m\in\mathcal{M}_1$, such that $i(x_1^{n_2}(m);y_2^{n_2})>\log(\textsfM_1)$, where $i(.;.)$ is the information density. If a unique index $m$ is found, set $\hat{m}_1=m$. Otherwise, it declares an error. Based on $\hat{m}_1$, find a unique $m\in\mathcal{M}_2$, such that $i(x_2^{n_2}(m);\tilde{y}_2^{n_2})>\log(\textsfM_2)$, where $\tilde{y}_2^{n_2}$ is the received signal without the signal $x_1^{n_2}$. If a unique index $m$ is found, set $\hat{m}_2=m$. Otherwise, it declares an error. At the weaker user (user 1), it finds a unique $m\in\mathcal{M}_1$, such that $i(x_1^{n_1}(m);y_1^{n_1})>\log(\textsfM_1)$. If a unique index $m$ is found, set $\hat{m}_1=m$.

\begin{rmk}
  The reason for considering the i.i.d. Gaussian codes instead of shell codes \cite{Polyanskiy_finite_block_length} in this work is as follows. Assume $x_1^{n_1}$ and $x_2^{n_2}$ are two shell codes. When user 2 tries to decode $m_1$ from the received signal $y_2^{n_2}=\sqrt{h_1}(x_1^{n_2}+x_2^{n_2})+z_{2}^{n_2}$, however, the truncated $x_1^{n_2}:=[x_{1,1},\ldots,\,x_{1,n_2}]$ is no longer a shell code. Then results in \cite{Tan_MUC_dispersion} and \cite{Sheldon_GBC_FBL}, which are based on shell codes, cannot be applied here. To avoid this issue, we apply i.i.d. Gaussian codes as a starting point to investigate the ED.
\end{rmk}


\subsection{Preliminaries}
When $n_1=n_2$, user 2 can perform the traditional two-step SIC. The first step is decoding user 1's codeword and removing its interference. Next, user 2 decodes his own codeword.
However, with $n_1>n_2$, the performance of using SIC is not clear. This motivates our investigation of using the ED as the first step of SIC. We first define a successful ED as follows.
\begin{definition}
A successful ED means that the user with a shorter blocklength constraint (user 2) can decode message of the user with the longer blocklength constraint (user 1) from the first $\tilde{n}_1$ received symbols: $Y_{2,1},\,Y_{2,2},\ldots,Y_{2,\tilde{n}_1}$, where $\tilde{n}_1\leq n_2<n_1$, while the resulting error probability fulfills the error probability constraint.
\end{definition}

Note that the channel output distribution of \eqref{EQ_BC_1st_ch_1} is still jointly Gaussian when $n_2<n_1$, if an i.i.d. Gaussian codebook is used. From \cite{Polyanskiy_finite_block_length} we know that the second-order achievable number of messages of a point-to-point channel with channel gain $h$, blocklength $n$, error probability constraint $\epsilon$, and power constraint $\textsfP$ with i.i.d. Gaussian input, can be specialized from our model by nulling $m_2$. It is described as follows:
\begin{align}\label{EQ_sub_opt_3rd_order_rate_G}
& \log \textsfM\leq  n{\normalfont\textsfC}(h\bar{\textsfP})-\sqrt{n\textsfV_G(h\cdot \bar{\textsfP})}Q^{-1}(\epsilon)+o\left(1\right),
\end{align}
where $\textsfV_G(\bar{\textsfP}):=\log^2 e\cdot \frac{\bar{\textsfP}}{1+\bar{\textsfP}}$ and $\bar{\textsfP}:=\textsfP-\delta,\,\delta>0$. Based on \eqref{EQ_sub_opt_3rd_order_rate_G}, we define the parameterized second order achievable rate as follows:
\begin{align}\label{EQ_parameterized_2nd_order_rate}
\textsfR(n,\SNR,\eps)&:=\textsfC(\SNR)-\sqrt{\frac{\textsfVG(\SNR)}{n}}Q^{-1}(\eps).
\end{align}

\section{Main Results}\label{Sec_MainResult}
In this section we introduce our main results: the necessary number of received symbols for a successful ED at the stronger user and the second-order achievable region $(\textsfM_1,\textsfM_2)$ of the two users under both IPC and SPC.

\subsection{Second Order Achievable Rate Region $(\textsfM_1,\textsfM_2)$ of ED with IPC}\label{Sec_Applications}

In the following, we apply the concept of ED to a two-user GBC with SIC.

\begin{Theorem}\label{Th_GBC_ED}
Denote the necessary number of symbols to successfully early decode user 1's signal at user 2 by $\tilde{n}_1$. Assume $n_2\leq n_1$ and $h_1\leq h_2$. If all the following conditions
\begin{align}
n_2&\geq \tilde{n}_1\geq\frac{\log\,{\normalfont\textsf{M}_1}}{\textsfC(g_2{\normalfont\bar{\textsf{P}}_1})-\log e\cdot \frac{g_2{\normalfont\bar{\textsf{P}}_1}}{2(1+g_2{\normalfont\bar{\textsf{P}}_1})}}+\frac{\log e \sqrt{4g_2{\normalfont\bar{\textsf{P}}_1}+2(g_2{\normalfont\bar{\textsf{P}}_1})^2} Q^{-1}({{\epsilon}_{SIC1}})}{2(1+g_2{\normalfont\bar{\textsf{P}}_1})\textsfC(g_2{\normalfont\bar{\textsf{P}}_1})-\log e\cdot g_2{\normalfont{\normalfont\bar{\textsf{P}}_1}}}\cdot\sqrt{n_1},\label{EQ_SIC_constraintED_ED_rate_improvement}
\end{align}
and
\begin{align}
\epsilon_{SIC1}+\epsilon_{SIC2}-\epsilon_{SIC1}\epsilon_{SIC2}+\epsilon_{1}\leq \epsilon, \,\,\quad 0< \eps_{1},\,\eps_{2},\,\eps_{SIC1},\,\eps_{SIC2}< 1 \label{EQ_Pe_constr}
\end{align}
hold, then under IPC all tuples of message sizes  $(\textsfM_1,\textsfM_2)$ in $\mathcal{M}_{IPC}$ are achievable, where
\begin{align}
\mathcal{M}_{IPC}=\Bigg\{(\textsfM_1,\textsfM_2): \,\,\log\textsfM_{1}&\leq n_1\bar{\textsfC}_1-\sqrt{n_1\bar{\textsfV}_1}Q^{-1}(\epsilon_1)+\mathcal{O}(1),\label{EQ_R1_ED}\\
\log\textsfM_{2}&\leq n_2\textsfC(h_2{\normalfont\bar{\textsf{P}}_2})-\sqrt{n_2{\normalfont \textsfV_G(h_2\bar{\textsf{P}}_2)}}Q^{-1}({\epsilon}_{SIC2})+\mathcal{O}(1)\Bigg\},\label{EQ_SIC_constraintED_R2_rate_improvement}
\end{align}
\begin{align}
\bar{\textsfC}_1&:=p\textsfC(g_1\bar{\textsfP}_1)+(1-p)\textsfC(h_1\bar{\textsfP}_1),\label{EQ_def_bar_C1}\\
\bar{\textsfV}_1&:=\log^2 e\cdot \left\{p\frac{g_1\bar{\textsfP}_1}{1+g_1\bar{\textsfP}_1}+(1-p)\frac{h_1\bar{\textsfP}_1}{1+h_1\bar{\textsfP}_1}\right\},\label{EQ_def_bar_V1}\\
g_1&:=\frac{h_1}{1+h_1\bar{\textsfP}_2},\label{EQ_def_g1}
\end{align}
$p:=\frac{n_2}{n_1},\,\normalfont\bar{\textsf{P}}_k:={\textsf{P}}_k-\delta,\,k=1,\,2,\,\delta>0$, ${\epsilon}_{SIC1}$ and ${\epsilon}_{SIC2}$ are the target decoding error probabilities of $m_1$ and $m_2$ at user 2 at the 1st and 2nd steps of SIC, respectively, $\normalfont g_2:=\frac{h_2}{1+h_2\bar{\textsf{P}}_2}$, and $g_2\normalfont\bar{\textsf{P}}_1$ is the equivalent output SNR at user 2 in the first step of SIC.
\end{Theorem}

The proof of the lower bound of $\tilde{n}_1$ is relegated to \ref{APP_ED_Gaussian_max_power} and the proof of the region $\mathcal{M}_{IPC}$ is relegated to \ref{APP_Pf_rate_region}.

In contrast to the pure private message case, the ED scheme can be naturally applied to a two-user GBC with only common-message (or equivalently, a multi-cast channel) and different decoding latencies. That is, a common message $m$ is sent, and the signal received at time $i$ at user $k$ is
$Y_{k,i}=\sqrt{h_k}X_i+Z_{k,i}, \,i=1,\ldots,n_k$, where $X_i\sim\mathcal{N}(0,\bar{\textsfP})$ is the transmitted symbol at time $i$. User $k$ wants to decode $m$ through $n_k$ received symbols.
We can compare this specialization to the asymptotic case. Denote the feasible lower bound of $\tilde{n}_1$ derived from the common-message specialization based one Theorem 1 by $g(n_1)$. Then we have the following
\begin{align}
\lim_{n_1\rightarrow\infty}\frac{\tilde{n}_1}{n_1}&\geq \lim_{n_1\rightarrow\infty}\frac{g(n_1)}{n_1}= \frac{\textsfC(h_1\bar{\textsfP})}{\textsfC(h_2\bar{\textsfP})-\frac{\log e\cdot h_2\bar{\textsfP}}{2(1+h_2\bar{\textsfP})}}\label{EQ_Azarian2_0}\\
&> \frac{\textsfC(h_1\bar{\textsfP})}{\textsfC(h_2\bar{\textsfP})},\label{EQ_Azarian2}
\end{align}
where \eqref{EQ_Azarian2} is the asymptotic result for the ED \cite{Azarian_ED}. The strict inequality comes from an strictly upper bounding during the error analysis shown in \ref{APP_ED_Gaussian_max_power}.

In contrast, without using ED or when \eqref{EQ_SIC_constraintED_ED_rate_improvement} is violated, TIN can be used, instead. Then user 2's second-order achievable $\log\textsfM_2$ can be derived from \eqref{EQ_sub_opt_3rd_order_rate_G} as follows
\begin{align}
\log\textsfM_2\leq n_2\textsfC\left(g_2\bar{\textsfP}_2\right)-\sqrt{n_2\textsfVG\left(\tilde{h}_2\bar{\textsfP}_2\right)}Q^{-1}(\epsilon_2)+\mathcal{O}(1),
\label{EQ_SIC_constraint_noED_R2_rate_improvement}
\end{align}
where $g_2:=\frac{h_2}{1+h_2\textsfP_1}$ and $\epsilon_2$ is user 2's target error probability.


\subsection{Rate region of ED with SPC}
The SPC with heterogenous blocklength constraints is stated as follows
\begin{align}\label{EQ_SPC_original_def}
\sum_{j=1}^{n_2}(x_{1,j}(m_1)+x_{2,j}(m_2))^2+\sum_{j=n_2+1}^{n_1}x_{1,j}^2(m_1)\leq n_1\textsfP_T,
\end{align}
for all $m_k\in\mathcal{M}_k,\,k=1,\,2.$ Because the cross-term in the first term on the left hand side (LHS) of \eqref{EQ_SPC_original_def} complicates the power allocation, we consider the following sum power constraint instead, for the following derivation and simulation:
\begin{align}\label{EQ_SPC_new_def}
\sum_{j=1}^{n_2}\left(x_{1,j}^2(m_1)+x_{2,j}^2(m_2)\right)+\sum_{j=n_2+1}^{n_1}x_{1,j}^2(m_1)\leq n_1\textsfP_T,
\end{align}
for all $m_k\in\mathcal{M}_k,\,k=1,\,2$. The validity of considering \eqref{EQ_SPC_new_def} instead of \eqref{EQ_SPC_original_def} is derived in Lemma \ref{Lemma_SPC_violation_prob} as shown below. We first assume
\begin{align}\label{EQ_SPC_assumption}
\sum_{j=1}^{n_2}x_{1,j}^2(m_1)\leq n_2\textsfP_{11},\, \sum_{j=1}^{n_2}x_{2,j}^2(m_2)\leq n_2\textsfP_{2}, \, \sum_{j=n_2+1}^{n_1}x_{1,j}^2(m_1)\leq (n_1-n_2)\textsfP_{12},
\end{align}
for all $m_k\in\mathcal{M}_k,\,k=1,\,2.$ Then we can consider the following power constraint instead of \eqref{EQ_SPC_original_def}
\begin{align}\label{EQ_SPC_new}
n_2(\textsfP_{11}+\textsfP_2)+(n_1-n_2)\textsfP_{12}\leq n_1\textsfP_T.
\end{align}
In short, if \eqref{EQ_SPC_assumption} and \eqref{EQ_SPC_new} are fulfilled, then the probability that the constraint in \eqref{EQ_SPC_original_def} is violated, is upper bounded by $e^{-\mathcal{O}(n_2)}$. Note that without incurring confusion, in the following we omit the parameterized $m_1$ and $m_2$ in codewords to simplify the notation.

\begin{lemma}\label{Lemma_SPC_violation_prob}
  Let $X_{1,j}\sim\mathcal{N}(0,\bar{\textsfP}_{11})\mbox{ and }\,X_{2,j}\sim\mathcal{N}(0,\bar{\textsfP}_{2})\mbox{ be i.i.d. generated and}\,\{X_{1,j}\}$ and $\{X_{2,j}\}$ are mutually independent, $ j=1,\cdots,n_2$, also let $X_{1,j}\sim\mathcal{N}(0,\bar{\textsfP}_{12})\mbox{ be i.i.d. generated},$ $j=n_2+1,\cdots,n_1$, where $\bar{\textsfP}_{11}=\textsfP_{11}-\delta$, $\bar{\textsfP}_{12}=\textsfP_{12}-\delta$, and $\bar{\textsfP}_{2}=\textsfP_{2}-\delta$, $\delta>0$. Then
  \begin{align}\label{EQ_cross_Pout}
  \Pr\left(\sum_{j=1}^{n_2}\left(X_{1,j}+X_{2,j}\right)^2+\sum_{j=n_2+1}^{n_1}X_{1,j}^2>n_1\textsfP_T\right)\leq e^{-\mathcal{O}(n_2)}
  \end{align}
  and
  \begin{align}\label{EQ_cross_Pout2}
  \Pr\left(\sum_{j=1}^{n_2}\left(X_{1,j}^2+X_{2,j}^2\right)+\sum_{j=n_2+1}^{n_1}X_{1,j}^2>n_1\textsfP_T\right)\leq e^{-\mathcal{O}(n_2)}.
  \end{align}
\end{lemma}

The proof is relegated to \ref{APP_Pr_cross_term_violation}. The main idea is to treat the event $\left\{\sum_{j=1}^{n_2}x_{1,j}x_{2,j}\geq n_2\delta\right\}$ as an outage and collect the probability of input violation ${\normalfont\mbox{Pr}}(\sum_{j=1}^{n_2}X_{1,j}X_{2,j}\geq n_2\delta)$ into the big-O term during the error analysis. After that, we use the concept of power backoff to ensure that by selecting $2n_2\delta$ as the power backoff, the total energy that is allocated to $\sum_{j=1}^{n_1}x_{1,j}^2$ and $\sum_{j=1}^{n_2}x_{2,j}^2$ will be no larger than $n_1P_T-2n_2\delta$ with probability close to 1. Based on Lemma \ref{Lemma_SPC_violation_prob}, we can extend the analysis in Theorem \ref{Th_GBC_ED} to the case with SPC as the following result.


\begin{Prop}\label{Lemma_HNOMA_sum_power}
Denote a power backoff for the sum power constraint \eqref{EQ_SPC_new_def} by $\delta$, such that $\bar{\textsfP}_{11}=\textsfP_{11}-\delta\geq 0$, $\bar{\textsfP}_{12}=\textsfP_{12}-\delta\geq 0$, $\bar{\textsfP}_{2}=\textsfP_{2}-\delta\geq 0$ fulfill
\begin{align}\label{EQ_SPC_power_backoff}
p\bar{\textsfP}_{11}+(1-p)\bar{\textsfP}_{12}\leq \textsfP_T-p\bar{\textsfP}_2.
\end{align}
Assume the blocklengths $n_1> n_2$ and channel gains $h_2>h_1$. If \eqref{EQ_SIC_constraintED_ED_rate_improvement}, \eqref{EQ_Pe_constr}, and \eqref{EQ_SPC_power_backoff} hold,
then under SPC all tuples of message sizes  $(\textsfM_1',\textsfM_2)$ in $\mathcal{M}_{SPC}$ are achievable, where
\begin{align}
\mathcal{M}_{SPC}=\Bigg\{(\textsfM_1',\textsfM_2): \,\,\log\textsfM_{1}'&\leq n_1\bar{\textsfC}'_1-\sqrt{n_1\bar{\textsfV}_1'}Q^{-1}(\epsilon_1)+\mathcal{O}(1),\label{EQ_R1_ED_p}\\
\log\textsfM_{2}&\leq n_2\textsfC(h_2{\normalfont\bar{\textsf{P}}_2})-\sqrt{n_2{\normalfont \textsfV_G(h_2\bar{\textsf{P}}_2)}}Q^{-1}({\epsilon}_{SIC2})+\mathcal{O}(1)\Bigg\},\label{EQ_SIC_constraintED_R2_rate_improvement}
\end{align}
and
\begin{align}
\bar{\textsfC}_1'&:=p\textsfC(g_1'\bar{\textsfP}_{11})+(1-p)\textsfC(h_1\bar{\textsfP}_{12}),\label{EQ_def_bar_C1_p}\\
\bar{\textsfV}_1'&:=\log^2e\cdot\left\{p\frac{g_1'\bar{\textsfP}_{11}}{1+g_1'\bar{\textsfP}_{11}}+(1-p)\frac{h_1\bar{\textsfP}_{12}}{1+h_1\bar{\textsfP}_{12}}\right\},\label{EQ_def_bar_V1_p}\\
g_1'&:=\frac{h_1}{1+h_1\bar{\textsfP}_2}.
\end{align}
\end{Prop}

The proof is relegated in \ref{APP_HNOMA_Sum_Power}.

\section{Numerical Results}\label{Sec_simulation}
In this section, we first formulate the achievable regions of the ED and HNOMA schemes as optimization problems, where the latter case is used as a comparison baseline. Based on these problem formulations, we then show the latency reduction of ED compared to the normal decoding, followed by the comparisons of achievable regions.

\subsection{Programming Formulation of the Achievable Rate Region}
Denote the target system error probability by $\epsilon$. Denote the rates at users 1 and 2 by $\textsfR_1$ and $\textsfR_2$, respectively. Let the intermediate variables $\eps_{SIC_k},\,k=1,\,2,$ denote the target error probabilities of decoding user $k$'s messages at the $k$-th step of SIC at user 2, respectively.
Let $\eps_{1}^{ED}$ denote the target error probability of decoding user 1's messages at user 1 when ED is used at receiver 2. Let $\eps_{1,j}^{HNOMA},\,j=1,\,2$ denote the target error probabilities of the sub-blocks 1 and 2, respectively, when HNOMA is used \cite{xu_hybrid_NOMA}. Note that for HNOMA, sub-blocks 1 and 2 are codewords with blocklengths $n_2$ and $n_1-n_2$, respectively. Denote the weighting in the weighted sum rate formulation by $\omega$, $0\leq\omega\leq 1$. Note that to optimize the sum-rate and rate region, the inequalities in the error probability constraints should be equalities, due to the tradeoff between the error probability and the rate in finite blocklength analysis.
Define
\begin{align}\label{EQ_Def_SNR}
 \SNR_{11}&:=\frac{h_1\bar{\textsfP}_1}{1+h_1\bar{\textsfP}_2},\quad\SNR_{21}:=\frac{h_2\bar{\textsfP}_1}{1+h_2\bar{\textsfP}_2},\quad\SNR_{12}:=h_1\bar{\textsfP}_1,\quad\mbox{and }\SNR_{22}:=h_2\bar{\textsfP}_2.
\end{align}

We then formulate an optimization problem for an enhanced HNOMA as follows:\\
$\mathbf{P_1^{IPC}}$ (enhanced weighted sum-rate of HNOMA with IPC):
\begin{align}
  \max &\quad \omega\textsfR_1\left(\eps_{1,1}^{HNOMA},\,\eps_{1,2}^{HNOMA}\right)+(1-\omega)\textsfR\left(n_2,\SNR_{22},\eps_{SIC2}\right)\label{EQ_P1_IPC_obj}\\
  \mbox{s.t.}&\quad 2-(1-\eps_{1,1}^{HNOMA})(1-\eps_{1,2}^{HNOMA})-(1-\eps_{SIC1})(1-\eps_{SIC2})\leq \epsilon\label{EQ_P1p_1st_constraint}\\
  &\quad 0< \eps_{SIC1},\,\eps_{SIC2},\,\eps_{1,1}^{HNOMA},\,\eps_{1,2}^{HNOMA}< 1,\label{EQ_P1P_IPC_eps}
\end{align}
where
\begin{align}
&\textsfR_1\left(\eps_{1,1}^{HNOMA},\,\eps_{1,2}^{HNOMA}\right)\notag\\
&:=p\cdot\min\left\{\textsfR(n_2,\,\SNR_{11},\,\eps_{1,1}^{HNOMA}),\,\textsfR(n_2,\,\SNR_{21},\,\eps_{SIC1})\right\}+\left(1-p\right)\textsfR(n_1-n_2,\,\SNR_{12},\,\eps_{1,2}^{HNOMA}).\label{EQ_P1P_IPC_R1}
\end{align}

Recall that $(\SNR_{11}, \SNR_{12})$ and $p$ are defined in \eqref{EQ_Def_SNR} and below \eqref{EQ_def_g1}, respectively and $\textsfR$ on the RHS in \eqref{EQ_P1P_IPC_R1} is defined in \eqref{EQ_parameterized_2nd_order_rate}; the minimum in \eqref{EQ_P1P_IPC_R1} is to ensure that the first sub-block of the weaker user can be decoded at the stronger user with an error probability $\eps_{SIC1}$ within blocklength $n_2$. Note that the higher order terms are neglected under the assumption of a sufficiently large blocklength.

To optimize the weighted sum-rate of SIC with ED, we formulate the following problem:

$\mathbf{P_2^{IPC}}$ (weighted sum-rate of ED with IPC):
\begin{align}\label{EQ_ED_sum_rate}
  \max&\quad \omega\textsfR_{1,ED}\left(\eps_{1}\right)+(1-\omega)\textsfR\left(n_2,\SNR_{22},\eps_{SIC2}\right)\\
  \mbox{s.t.} &\quad \eqref{EQ_SIC_constraintED_ED_rate_improvement},\,\eqref{EQ_Pe_constr}, \notag\\
 &\quad n_2\geq\frac{n_1\textsfR_{1,ED}(\eps_1)}{\textsfC(\SNR_{21})-\log e\cdot \frac{\SNR_{21}}{2(1+\SNR_{21})}}+\frac{\log e \sqrt{4\SNR_{21}+2\SNR_{21}^2} Q^{-1}(\eps_{SIC1})}{2(1+\SNR_{21})\textsfC(\SNR_{21})-\log e\cdot \SNR_{21}}\cdot\sqrt{n_1},\label{EQ_Cond_ED}
\end{align}
where $\textsfR_{1,ED}$ is derived from \eqref{EQ_R1_ED} as follows:
\begin{align}
\textsfR_{1,ED}\left(\eps_{1}\right)&:=\textsfbarC_1-\sqrt{\frac{\textsfbarV_1}{n_1}}Q^{-1}(\eps_{1}),\label{EQ_P2_R1}
\end{align}
and $\bar{\textsfC}_1:=p\textsfC(g_1\bar{\textsfP}_1)+(1-p)\textsfC(h_1\bar{\textsfP}_1),\,
\bar{\textsfV}_1:=\log e^2\cdot \left\{p\frac{g_1\bar{\textsfP}_1}{1+g_1\bar{\textsfP}_1}+(1-p)\frac{h_1\bar{\textsfP}_1}{1+h_1\bar{\textsfP}_1}\right\},\,
\mbox{ and }g_1:=\frac{h_1}{1+h_1\bar{\textsfP}_2}.$

Note that \eqref{EQ_Cond_ED} ensures that the weaker user's signal can be decoded with error probability $\eps_{SIC1}$ within blocklength $n_2$ and the channel SNR is as $\SNR_{21}$, which plays a similar role to the minimum in \eqref{EQ_P1P_IPC_R1} of the HNOMA. Note also that we assume that the blocklength is sufficiently large, such that we omit the big-O term in the programming formulation.

For the SPC, we first redefine $\SNR_{11}:=\frac{h_1\bar{\textsfP}_{11}}{1+h_1\bar{\textsfP}_2},\,\quad\SNR_{12}:=h_1\bar{\textsfP}_{12},\,\quad \mbox{and }\SNR_{21}:=\frac{h_2\bar{\textsfP}_{11}}{1+h_2\bar{\textsfP}_2}$. Then we formulate the following optimization problem for the enhanced weighted sum-rate of HNOMA:\\
$\mathbf{P_1^{SPC}}$ (enhanced weighted sum-rate of HNOMA with SPC):
\begin{align}\label{EQ_HNOMA_sum_rate_SPC}
  \max &\quad \omega\textsfR_1\left(\eps_{1,1}^{HNOMA},\,\eps_{1,2}^{HNOMA}\right)+(1-\omega)\textsfR\left(n_2,\SNR_{22},\eps_{SIC2}\right) \\
  \mbox{s.t.} &\quad \bar{\textsfP}_{11}\geq 0,\,\bar{\textsfP}_{12}\geq 0,\, \bar{\textsfP}_{2}\geq 0,\label{EQ_constraint_positive_power}\\
  &\quad \eqref{EQ_SPC_power_backoff},\,\eqref{EQ_P1p_1st_constraint},\,\eqref{EQ_P1P_IPC_eps}.\notag
\end{align}

The optimization problem of the ED with SPC is formulated as follows:

$\mathbf{P_2^{SPC}}$ (weighted sum-rate of ED with SPC):
\begin{align}
  \max&\quad \omega\textsfR'_{1,ED}\left(\eps_{1}\right)+(1-\omega)\textsfR\left(n_2,\SNR_{22},\eps_{SIC2}\right)\\
  \mbox{s.t.}&\quad \eqref{EQ_SIC_constraintED_ED_rate_improvement},\,\eqref{EQ_Pe_constr},\,\eqref{EQ_SPC_power_backoff},\,\eqref{EQ_constraint_positive_power},
\end{align}
where $\textsfR'_{1}\left(\eps_{1}^{ED}\right)$ is derived from \eqref{EQ_R1_ED_p}.

\subsection{Latency Reduction}{
We compare the latencies among three cases: 1) the number of received symbols necessary for a successful ED under asymptotic analysis; 2) decoding after the complete codeword is received (without ED); 3) the derived number of received symbols necessary for a successful ED under finite blocklength analysis. We consider the following setting: $\epsilon=2\cdot 10^{-6}$, $h_1=1$, $\textsfP_1=8\mbox{ and }\textsfP_2=0.2$ for IPC. We consider three different blocklengths: $n_1=$512, 1024, and 2048. The three schemes are compared in Fig. \ref{Fig_comparison_earlydecoding_max}. Without ED, the stronger user can start to decode only after receiving $n_1$ symbols. When ED is activated and successful, we can observe that the improvement of the latency reduction increases with an increasing $h_2$. We use \eqref{EQ_Azarian2}, which is a latency lower bound, as the baseline for comparison. When the channel gain is sufficiently large, the early decoding gain in terms of latency reduction can be achieved around 450 symbols when $n_1=2048$ is considered, while the gap of ED to the asymptotic scenario is around 350 symbols. Note that the gap between the results of the early decodings in asymptotic and finite blocklength analyses is not only from the channel dispersion but also from the bounding error when deriving \eqref{EQ_SIC_constraintED_ED_rate_improvement}.


\begin{figure}[h!]
\centering \epsfig{file=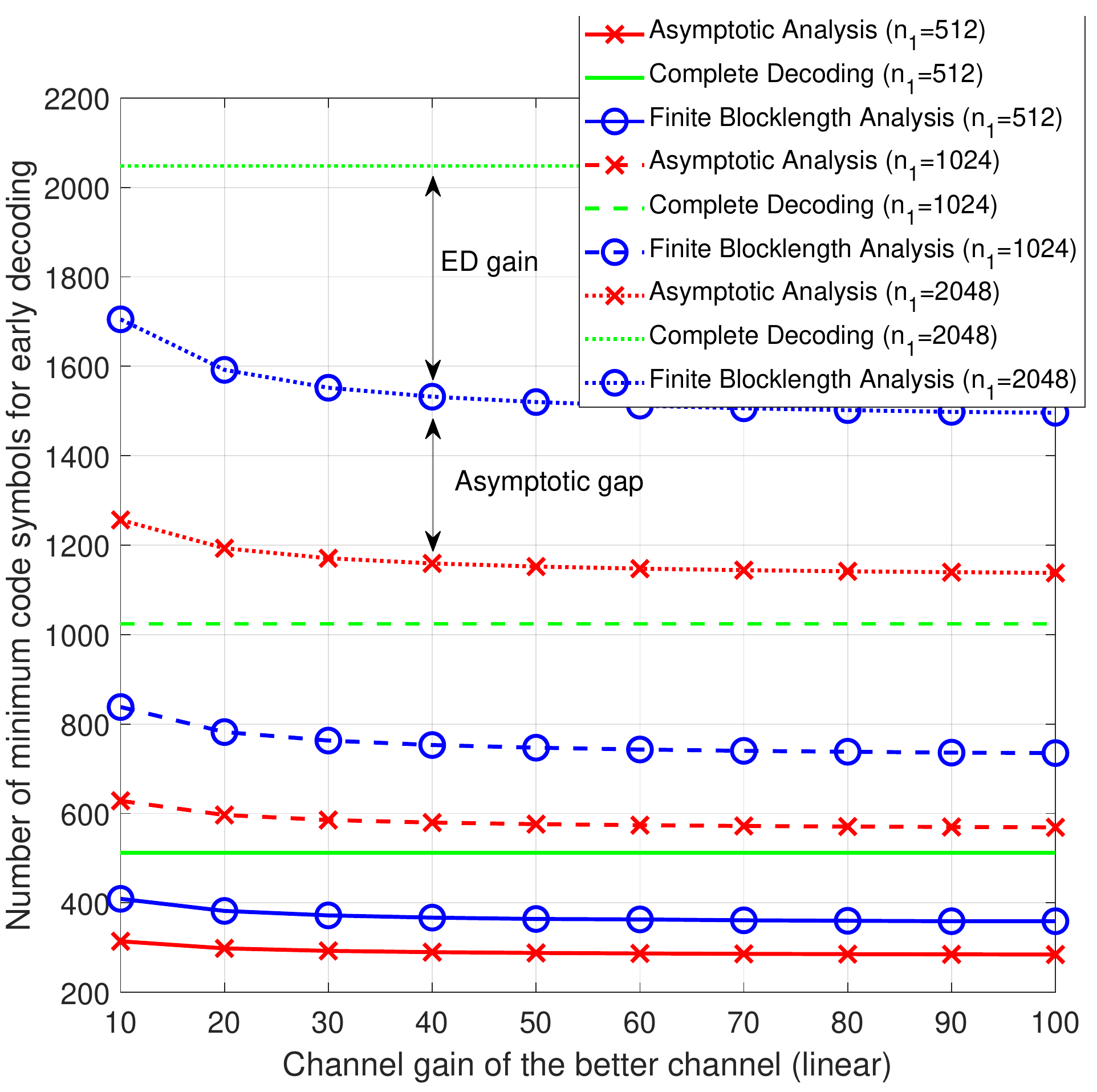, width=0.6\textwidth}
\caption{Comparison of necessary numbers of received symbols for successfully decoding a message under the constraints of average probability of error under maximal channel input power constraint: ED with finite blocklength,
ED with infinite blocklength, and complete decoding.}
\label{Fig_comparison_earlydecoding_max}
\end{figure}

Assume that we consider the case in which $n_2$ is set as the lower bound in \eqref{EQ_SIC_constraintED_ED_rate_improvement}. Then we compare \eqref{EQ_SIC_constraintED_R2_rate_improvement} and \eqref{EQ_SIC_constraint_noED_R2_rate_improvement} in Fig. \ref{Fig_comparison_SIC}, where the rate gain by using the ED over TIN is apparent. In particular, about 7-dB SNR gain can be achieved by the ED under our setting.
Please note that similar comparisons as Fig. \ref{Fig_comparison_earlydecoding_max} and Fig. \ref{Fig_comparison_SIC} for GBC with only common messages can be seen in \cite{PHL_ISIT21}.

\begin{figure}[h!]
\centering \epsfig{file=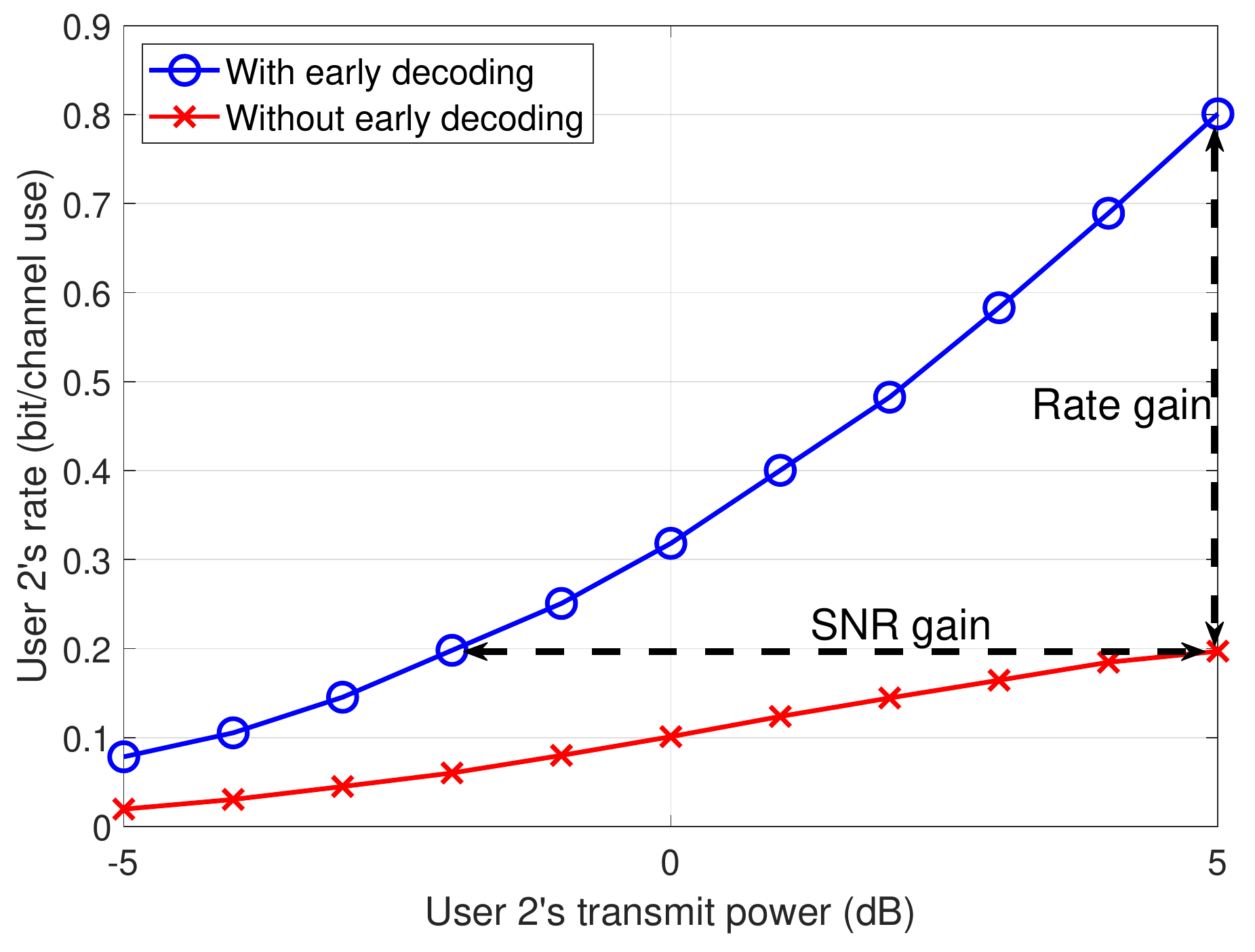, width=0.6\textwidth}
\caption{Comparison of the stronger users' rates with and without ED given a latency constraint.}
\label{Fig_comparison_SIC}
\end{figure}
}

\subsection{Comparison of Sum Rate/Rate Region}

\subsubsection{Sum-rate comparison}
\begin{figure}[h!]
\centering \epsfig{file=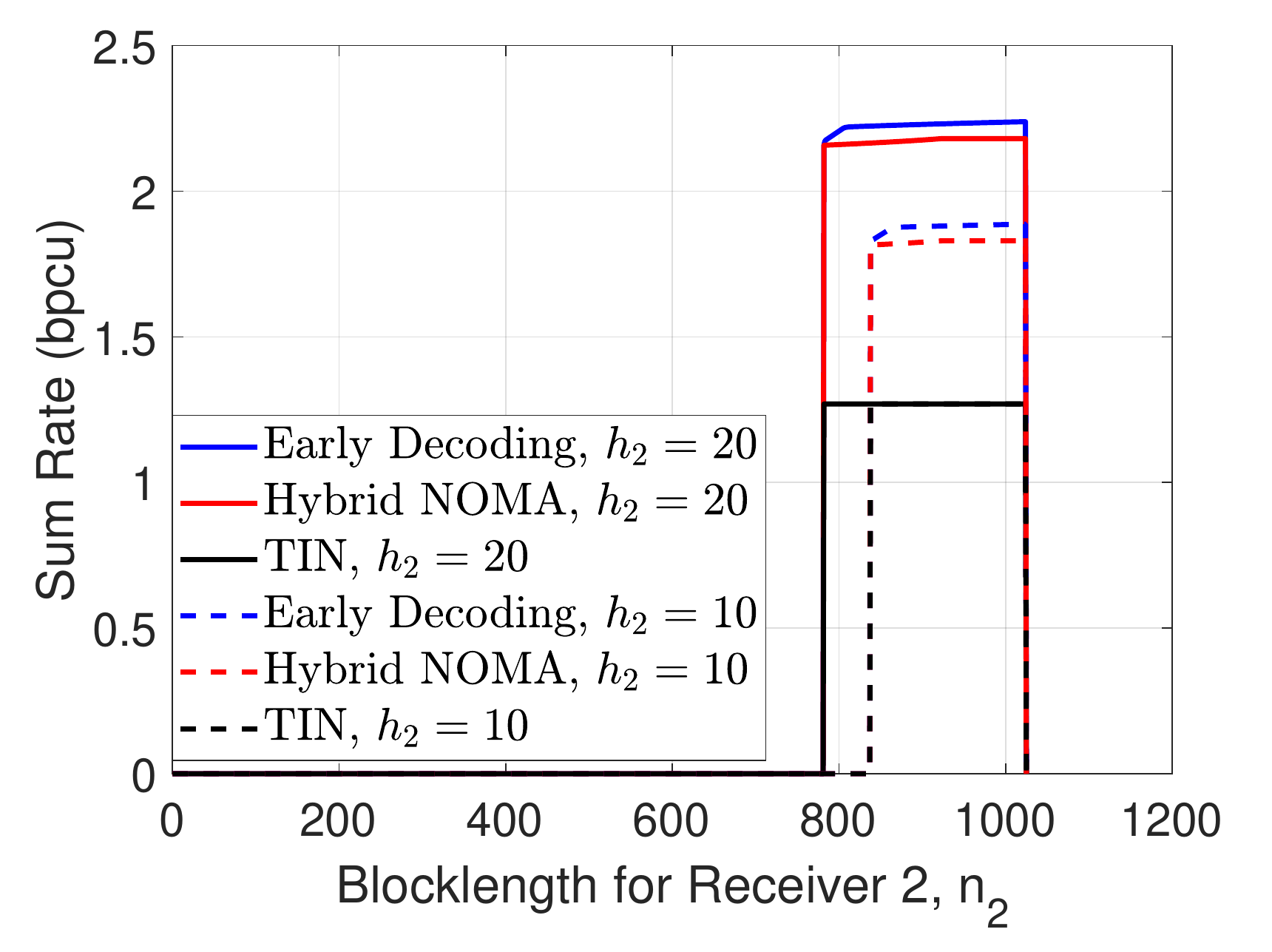, width=0.6\textwidth}
\caption{Comparison of the sum rates between ED and HNOMA under IPC with different $h_2$.}
\label{Fig_comparison_SR_IPC1}
\end{figure}

In the following, we use two examples to compare the sum-rates ($\omega=\frac{1}{2}$) of the ED in $\mathbf{P_2^{IPC}}$ and HNOMA in $\mathbf{P_1^{IPC}}$, under individual power constraints. We use grid search to find the optimal solution in the three problems with step sizes $\epsilon/100$. We only show the range of $n_2$ where $\mathbf{P_2^{IPC}}$ is feasible. Therefore, the sum rates below a threshold of $n_2$ will be zero. We consider the following setting:
$n_1=1024,\,h_1=1,\textsfP_1=8,\,\textsfP_2=0.2,\,\epsilon=2\cdot 10^{-6}$ under $h_2=10$ and $20$. We can find that ED outperforms HNOMA for all $n_2$ feasible for ED, under individual power constraint, for $h_2\in\{10,\,20\}$. In the same figure we can also observe that a larger $h_2$ will not only enlarge the feasible region of operating the ED but also enhance the sum-rate performance, which is consistent with the intuition. Besides, we can interpret curves in Fig. \ref{Fig_comparison_SR_IPC1} as a tradeoff between latency and sum-rate. In particular, when a lower latency at the stronger user is requested, we can use a smaller $n_2$ if the ED is feasible. However, due to the error probability constraint, the stronger user's rate will also be reduced, which causes the sum-rate as an increasing function of $n_2$.

\subsubsection{Rate regions comparison}
\begin{figure}[htp]
    \centering \epsfig{file=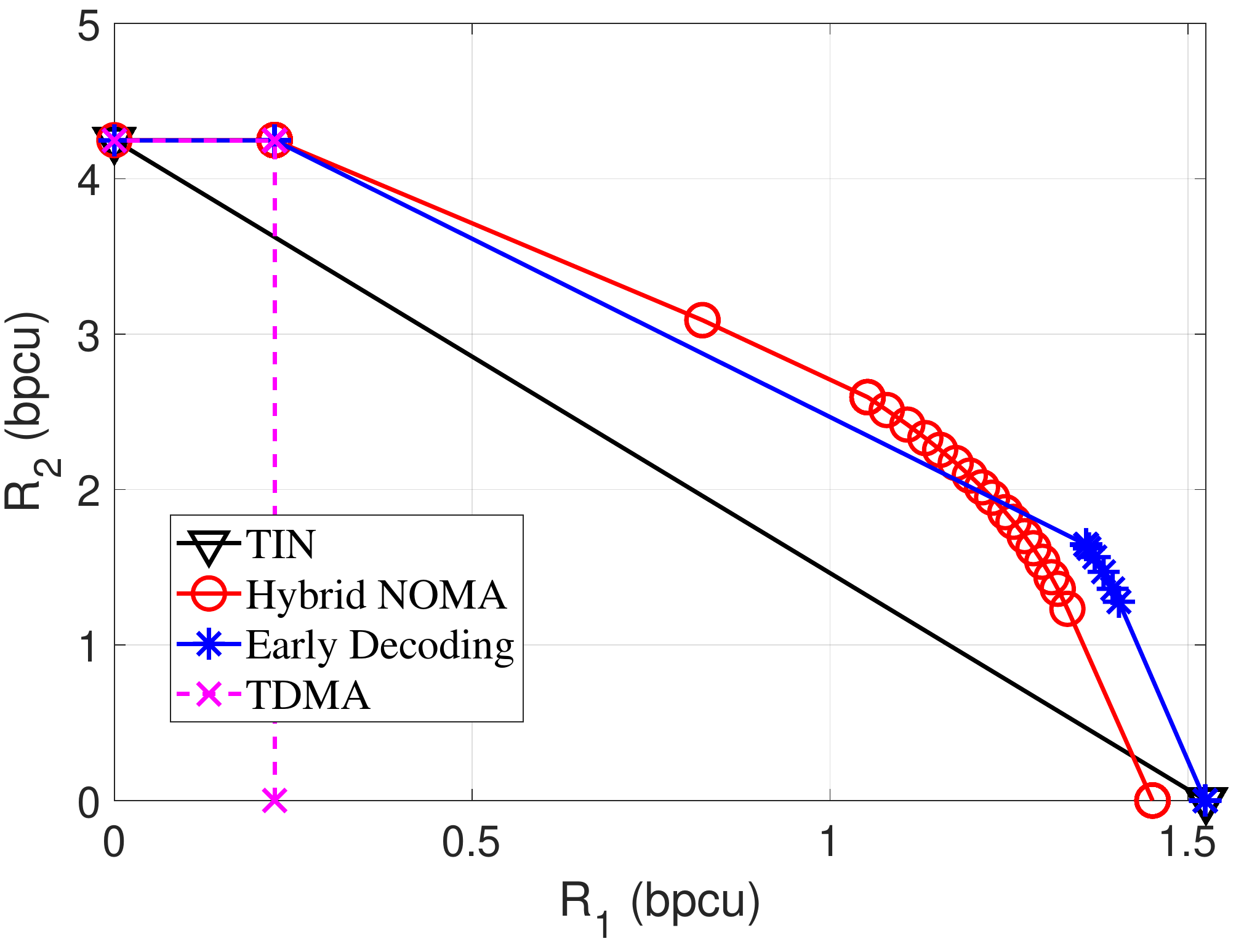, width=0.6\textwidth}
    \caption{Comparison of achievable rate regions: $h_1=1, h_2=50, P_T=10, n_1 = 1024, n_2=840$.}
    \label{rate_region_50}
\end{figure}

We now consider the rate regions under SPC from solving $\mathbf{P_1^{SPC}}$ and $\mathbf{P_2^{SPC}}$ with the following system parameters: $h_1=1, \,h_2=50, \,P_T=10, \,n_1 = 1024, \,n_2=840, \,\epsilon=2\cdot 10^{-5}$ in Fig. \ref{rate_region_50}. We can observe that both ED and HNOMA have their own advantages. As for ED, it can be beneficial when transmitting a single codeword with a longer blocklength compared to HNOMA, whose weaker user decodes two concatenated shorter codewords. Therefore, when the weighting $\omega$ is higher, ED outperforms HNOMA when $h_2$ is sufficiently large. On the contrary, by HNOMA, the transmission consists of 2 segments, which brings the flexibility of non-overlapping transmission, and therefore, it can outperform ED when $h_2$ is smaller. In particular, when $h_2$ is smaller, the feasibility constraint \eqref{EQ_SIC_constraintED_ED_rate_improvement} is harder to fulfill.

%

\section{conclusion}\label{Sec_conclusion}
We investigate a two-user Gaussian broadcast channel with heterogeneous blocklength constraints. Unlike the traditional GBC where two users have the same blocklength constraints, here, the user with higher output SNR has a shorter blocklength constraint. We show that with sufficiently large output SNR, the stronger user can perform early decoding to decode the interference, followed by the successive interference cancellation, which is not yet reported in the literature. To achieve this goal, we derive an explicit lower bound on the necessary number of received symbols for a successful ED, using an independent and identically distributed Gaussian input. A second-order rate of the weaker user who suffers from an SNR change due to the heterogeneous blocklength constraint, is also derived. Numerical results show that ED has a larger sum rate and rate region than HNOMA, when the channel gain of the better channel is sufficiently larger than the other one. Then ED with SIC is a promising technique for future broadcast channels with heterogeneous blocklength constraints. Our future works include deriving a second order capacity outer bound and improving the second order performance by a better input distribution than i.i.d. Gaussian input.

\renewcommand{\thesection}{Appendix I}
\section{Proof of Theorem \ref{Th_GBC_ED}}\label{APP_ED_Gaussian_max_power}
To derive the necessary number of received symbols, namely, $n_2$, for a successful ED, we investigate the error analysis at the stronger user. We consider the following average error probability (at the first step of SIC) at user 2, from the dependence testing bound given a specific code $\mathcal{C}$ \cite[Lemma 19]{Polyanskiy_finite_block_length}:
\begin{align}
&\frac{1}{\textsf{M}_1}\sum_{m_1=1}^{\textsf{M}_1}\mbox{Pr}(\hat{m}_1\neq m_1\mbox{ at user }2|\,\,m_1 \mbox{ is sent, } \mathcal{C} \mbox{ is used} )\notag\\
\leq &\frac{1}{\textsfM_1}\sum_{m=1}^{\textsfM_1} \left\{\mathds{1}_{x_1^{n_1}(m)\notin\mathcal{F}^{n_1}}+P_{Y_2 ^{{n_2} }|X_1^{{n_2} }=x_1^{n_2}(m)}\left(i(x_1^{n_2}(m);Y_2 ^{{n_2} })\leq \log \textsfM_1\right)\right.\notag\\
&\hspace{8cm}\left.+\textsfM_1 P_{Y_2 ^{{n_1} }}\left(i(x_1^{n_2}(m);Y_2 ^{{n_2} }) >\log \textsfM_1\right)\right\}\label{EQ_DT_formulation1}\\
\leq &\frac{1}{\textsfM_1}\sum_{m=1}^{\textsfM_1} \left\{\mathds{1}_{||x_1^{n_1}(m)||^2>n_1\textsfP_1}+\mathds{1}_{||x_1^{n_1}(m)||_{\infty}>{n_1}^{a}}+P_{Y_2 ^{n_2 }|X_1^{{n_2} }=x_1^{n_2}(m)}\left(i(x_1^{n_2}(m);Y_2 ^{{n_2} })\leq \log \textsfM_1\right)\right.\notag\\
&\hspace{8cm}\left.+\textsfM_1 P_{Y_2 ^{{n_2} }}\left(i(x_1^{n_2}(m);Y_2 ^{{n_2} }) >\log \textsfM_1\right)\right\}\label{EQ_DT_input_constraint_union}\\
\leq& \frac{1}{\textsfM_1}\sum_{m=1}^{\textsfM_1} \left\{P_{\tilde{Y}_2^{n_2}|X_1^{n_2}=x_1^{n_2}(m)}\left(i(x_1^{n_2}(m);\tilde{Y}_2^{n_2})\leq \log \textsfM_1\right)\right.\notag\\
&\hspace{4cm}\left.+\textsfM_1P_{\tilde{Y}_2^{n_2}}\left(i(x_1^{n_2}(m);\tilde{Y}_2^{n_2}) >\log \textsfM_1\right)\right\}+e^{-\frac{n_2\delta^2}{4}}+e^{\frac{-n_1^{2a}}{2\textsfP_1}+\ln (2n_1)},\label{EQ_Pe_max_power}
\end{align}
where the code $\mathcal{C}$ is specified by \eqref{EQ_sub_opt_3rd_order_rate_G}. In particular, from random coding analysis there must exist a code $\mathcal{C}=\{x_1^{n_1}(1),\,x_1^{n_1}(2),\,\ldots x_1^{n_1}(\textsfM_1)\}$ with blocklength ${n_1}$ achieving the rate \eqref{EQ_sub_opt_3rd_order_rate_G} while fulfilling the input power and error probability constraints, where each codeword is i.i.d. generated according to $\prod_{k=1}^{n_1} p_{X_1}(x_{1,k})$, $x_{1,k}\in\mathcal{X}_1,\,k=1,\cdots,n_1$. In \eqref{EQ_DT_formulation1}, $\mathcal{F}^{n_1}$ is the channel input constraint, including the maximal power constraint \eqref{EQ_max_power_constraint_BC} and the peak constraint $||x_1^{n_1}(m)||_{\infty}\leq {n_1}^{
a}$ \footnote{The peak constraint is only for the proof purpose, i.e., to ensure the vanishing property of the Berry-Esseen ratio, in particular, when calculating the third moment of the information density. Furthermore, with this additional constraint, the derived rate is a lower bound of the achievable rate of the original system.}, $0<a<1$. Then the first term in \eqref{EQ_DT_formulation1} (with the normalization with respect to $\textsfM_1$) is the probability of input-constraint violation, the third and the fourth terms are outage and confusion probabilities, respectively, given a specific codeword \cite[(75)]{Polyanskiy_finite_block_length}. With the selected code $\mathcal{C}$, we can re-map the messages of all the codewords violating the input constraint to one arbitrary vector which fulfills the power constraint while the decoding region is kept unchanged by this remapping. Under such a setting, the probability of the input power constraint being violated is merged into the decoding error probability \cite[Theorem 20]{Polyanskiy_finite_block_length}. In \eqref{EQ_Pe_max_power}, we use concentration inequalities to upper bound the violation probability of maximal power constraint and the peak constraint. A detailed derivation can be seen in \cite{PHL_arxiv21_J_ED}.

In the first step of SIC at user 2, the received signal can be equivalently expressed as
\begin{align}\label{}
  \tilde{Y}_2:=\sqrt{g_2}X_1+{\tilde{Z}}_2,
\end{align}
where ${\tilde{Z}}_2\sim\mathcal{N}(0,1),\,\tilde{Z}_2\ind X$, and
\begin{align}\label{EQ_eq_gain_strong_user}
  g_2:=\frac{h_2}{1+h_2\textsfP_2}.
\end{align}

%

Fix any codeword $x_1^{n_2}(m),\,m\in\{1,\ldots,\,\textsfM_1\}$ from $\mathcal{C}$, the information density $i(x_1^{n_2}(m); \tilde{Y}_2^{n_2})$ can be calculated as follows:
\begin{align}
{i(x_1^{n_2}(m); \tilde{Y}_2^{n_2})}
=&\log\frac{(2\pi)^{-{n_2}/2}e^{-\frac{||\tilde{Y}_2^{n_2}-\sqrt{g_2}x_1^{n_2}(m)||^2}{2}}}{(2\pi(1+g_2\textsfP_1))^{-{n_2}/2}
e^{-\frac{||\tilde{Y}_2^{n_2}||^2}{2(1+g_2\textsfP_1)}}}=\sum_{j=1}^{n_2}W_j,\label{EQ_info_density_decompose}
\end{align}
where the second equality in \eqref{EQ_info_density_decompose} is due to the memoryless channel, and we define
\begin{align}\label{EQ_Def_W}
  W_j:=\textsfC(g_2\textsfP_1)+\frac{\log e\cdot g_2(x_{1,j}^2-\textsfP_1\tilde{Z}_j^2)}{2(1+g_2\textsfP_1)}+\frac{\log e}{1+g_2\textsfP_1}\sqrt{g_2}x_{1,j}\tilde{Z}_j.
\end{align}
The mean of $W_j$ conditioned on $x_{1,j}$ is as follows
\begin{align}\label{EQ_info_density_mean}
  \mathds{E}_{{ \tilde{Y}_{2,j}}|X_{1,j}=x_{1,j}}[W_j]=\textsfC(g_2\textsfP_1)+\frac{\log e\cdot g_2(x_{1,j}^2-\textsfP_1)}{2(1+g_2\textsfP_1)}.
\end{align}

Then the centralized information density of the $j$-th symbol conditioned on $x_{1,j}$ is as follows:
\begin{align}\label{EQ_centralized_info_density}
W_j-\mathds{E}_{{ \tilde{Y}_{2,j}}|X_{1,j}=x_{1,j}}[W_j]
=&\frac{\log e}{1+g_2\textsfP_1}\left( \sqrt{g_2}x_{1,j}\tilde{Z}_{2,j}+g_2\frac{\textsfP_1}{2}(1-\tilde{Z}_{2,j}^2)\right).
\end{align}
To upper bound both the confusion (when the wrong codewords are treated as the transmitted ones) and outage (when the correct codewords is treated not be transmitted) probabilities, we derive the Berry-Esseen (B-E) ratio as follows. First, the absolute centralized third moment of the information density given $x_1^n$ can be upper bounded through standard analysis as follows
\begin{align}
&\hspace{-1cm}\sum_{j=1}^{n_2}\mathds{E}_{{ \tilde{Y}_{2,j}}|X_{1,j}=x_{1,j}}\left[\left|W_j-\mathds{E}_{{ \tilde{Y}_{2,j}}|X_{1,j}=x_{1,j}}[W_j]\right|^3\right]
\leq  4 \sum_{j=1}^{n_2}\left(\frac{\log e\cdot \sqrt{g_2}}{1+g_2\textsfP_1}\right)^3\left(8(\sqrt{g_2}\textsfP_1)^3+2|x_{1,j}|^3\right).\label{EQ_T4}
\end{align}


The variance of the information density given $x_1^{n_2}$ can be lower bounded as follows:
\begin{align}
\sum_{j=1}^{n_2}\mbox{Var}_{\tilde{Y}_{2,j}|X_{1,j}=x_{1,j}}\left[W_j\right]
&=\sum_{j=1}^{n_2}\mathds{E}_{\tilde{Y}_{2,j}|X_{1,j}=x_{1,j}}\left[\left(
\frac{\log e}{1+g_2\textsfP_1}\left( \sqrt{g_2}x_{1,j}\tilde{Z}_{2,j}+g_2\frac{\textsfP_1}{2}(1-\tilde{Z}_{2,j}^2)\right)\right)^2\right]\label{EQ_V0}\\
&=\sum_{j=1}^{n_2}\left(\frac{\log e}{1+g_2\textsfP_1}\right)^2g_2\left(x_{1,j}^2+g_2\frac{\textsfP_1^2}{2}\right)\geq {n_2}\left(\frac{\log e\cdot g_2\textsfP_1}{\sqrt{2}(1+g_2\textsfP_1)}\right)^2,\label{EQ_V}
\end{align}
where \eqref{EQ_V0} is from \eqref{EQ_centralized_info_density} and in \eqref{EQ_V}, we remove the term related to $x_{1,j}^2$.

After substituting \eqref{EQ_T4} and \eqref{EQ_V} into Berry-Esseen theorem, we can upper bound the Berry-Esseen ratio conditioned on $x_1^{n_2}$ as follows
\begin{align}
\frac{T}{V^{3/2}}&\leq \frac{4\sum_{j=1}^{n_2}\left(\frac{\sqrt{g_2}}{1+g_2 \textsfP_1}\right)^{3}\left(8({\sqrt{g_2}}\textsfP_1)^3+2|x_{1,j}|^3\right)}{\left(\frac{1}{2}{n_2}\left(\frac{\sqrt{g_2}\textsfP_1}{1+g_2 \textsfP_1}\right)^2\right)^{\frac{3}{2}}}\label{EQ_B_1}\\
&:= \frac{4d_1\left( 8n_2(\sqrt{g_2}\textsfP_1)^3+2\sum_{j=1}^{n_2}|x_{1,j}|^3 \right)}{\left(\frac{1}{2}d_1n_2\right)^{\frac{3}{2}}}\label{EQ_B1_1}\\
&\leq c_0\cdot {n_2}^{-\frac{1}{2}}+ c_1\cdot {n_2}^{a-\frac{1}{2}}:=B_0(n_2),\label{EQ_B_2}
\end{align}
where in \eqref{EQ_B1_1}, we define $d_1:=\left(\frac{ \sqrt{g_2}}{1+ g_2\textsfP_1}\right)^3$, which is a constant only dependent on $\textsfP_1$ and channel gains, but independent of $n_2$; in \eqref{EQ_B_2}, we define $c_0:=\frac{62\sqrt{2}(\sqrt{g_2}\textsfP_1)^3}{\sqrt{d_1}}$, $c_1:=\frac{128\sqrt{2}\textsfP_1}{\sqrt{d_1}p^{1+a}}$ and the inequality comes from the assumption of peak constraint $|x_{1,j}|\leq {n_1}^{a}$, $a<\frac{1}{2}$, with the following upper bounding
\begin{align}
\sum_{j=1}^{n_2} |x_{1,j}|^3 &\leq |\max\, x_{1,j}|\sum_{j=1}^{n_2} |x_{1,j}|^2\overset{(a)}{\leq} |\max\, x_{1,j}| {n_1}\textsfP_1\label{EQ_power3_2}\overset{(b)}{\leq} {n_1}^{a+1}\textsfP_1= \left(\frac{{n_2}}{p}\right)^{a+1}\textsfP_1,
\end{align}
where (a) comes from the maximal power constraint \eqref{EQ_max_power_constraint_BC} and (b) comes from the assumption of peak constraint.

Besides, the confusion probability conditioned on $x_1^{n_2}$ can be upper bounded as
\begin{align}
P_{\tilde{Y}_2^{n_2}}\left[\sum_{j=1}^{n_2}W_j>\log\gamma_{n_2}\right]&= \mathds{E}_{\tilde{Y}^{n_2}|X^n=x^n}\left[\exp\left(-\sum_{j=1}^{n_2}W_j\cdot\mathds{1}\left\{\frac{\sum_{j=1}^{n_2}W_j}{n_2}>\frac{\log \gamma_{n_2}}{n_2}\right\}\right)\right]\label{EQ_Yury_change_of_measure}\\
&\overset{(a)}{\leq} \frac{2}{\gamma_{n_2}}\left(\frac{\ln 2}{\sqrt{
\pi d_1 n_2}}+B_0(n_2)\right)\label{EQ_Q_Yn_xn0}\\
&:=\frac{B_1(n_2)}{\gamma_{n_2}}, \label{EQ_Q_Yn_xn}
\end{align}
where \eqref{EQ_Yury_change_of_measure} is from the change of measure \cite[(257)]{Polyanskiy_finite_block_length}, \eqref{EQ_Q_Yn_xn0} is from \cite[Lemma 47]{Polyanskiy_finite_block_length} with \eqref{EQ_V}, which is upper bounded by the same step used in \eqref{EQ_B_1}. In addition, we use \eqref{EQ_B_2} to bound the Berry-Esseen ratio. Note that $B_1(n_2)$ is a constant depending only on $n_2$, $h_1$, and $\textsfP_2$ but not on the realization $x^{n_2}$. Therefore, the total confusion probability can be simply derived from \eqref{EQ_Q_Yn_xn} as
\begin{align}
\textsfM_1\cdot P_{X^{n_2}}P_{\tilde{Y}_2^{n_2}}\left[\sum_{j=1}^{n_2}W_j>\log\gamma_{n_2}\right]&\leq \frac{\textsfM_1\cdot B_1(n_2)}{\gamma_{n_2}}.\label{EQ_total_P_conf}
\end{align}

Meanwhile, again by Berry-Esseen theorem, the outage probability conditioned on $ {x}^{n_2}$ can be expressed as follows
\begin{align}
P_{\tilde{Y}_2^{n_2}|X_1^{n_2}=x_1^{n_2}}\left[\sum_{j=1}^{n_2}W_j\leq \log \gamma_{n_2}\right]&\leq Q\left(r_m({n_2})\right)+B_0(n_2),\label{EQ_P_out}
\end{align}
where
\begin{align}\label{EQ_Def_r}
r_m({n_2}):=\frac{n_2\mu_m-\log \textsfM_1}{n_2\sigma_m},
\end{align}
$\mu_m$ and $\sigma^2_m$ are defined as the RHS of \eqref{EQ_info_density_mean} and the LHS of \eqref{EQ_V}, respectively, and $B_0(n_2)$ is defined in \eqref{EQ_B_2}.

By selecting $\gamma_{n_2}=\textsfM_1$, we can then bound the conditional confusion and outage probabilities in \eqref{EQ_total_P_conf} and \eqref{EQ_P_out} respectively as follows
\begin{align}
\textsfM_1\cdot P_{X_1^{n_2}}P_{\tilde{Y}_2^{n_2}}\left[\sum_{j=1}^{n_2}W_j>\log\gamma_{n_2}\right]&\leq B_1(n_2),\label{EQ_confusion_prob_2}\\
P_{\tilde{Y}_2^{n_2}|X_1^{n_2}=x_1^{n_2}(m)}\left(\sum_{j=1}^{n_2}W_j\leq \log \textsfM_1\right)
&=  Q\left(r_m({n_2})\right)+B_0(n_2).\label{EQ_UB_given_x}
\end{align}
Note that we need $r_m({n_2})>0$ since we consider the case in which $\epsilon_{SIC1}<\frac{1}{2}$. Note also that $r_m({n_2})$ is a function of $||x_1^{n_2}(m)||^2$. To derive an upper bound of \eqref{EQ_UB_given_x}, we resort to finding a lower bound of $r_m({n_2})$ since $Q$-function is monotonically decreasing. Furthermore, we aim to find a uniform lower bound of $r_m({n_2})$, which will be independent of the given $x_1^{n_2}(m)$, as shown as follows
\begin{align}
r_m({n_2})=&\frac{n_2\textsfC(g_2\textsfP_1)+\frac{\log e\cdot g_2}{2(1+g_2\textsfP_1)}\left(||x_1^{n_2}(m)||^2-n_2\textsfP_1\right)-\log \textsfM_1}{\frac{\log e}{2(1+g_2\textsfP_1)}\sqrt{4g_2||x_1^{n_2}(m)||^2+2{n_2}g_2^2\textsfP_1^2}}\label{EQ_r_intermediate2_0}\\
\geq &\frac{2(1+g_2\textsfP_1)\left({n_2}\textsfC(g_2\textsfP_1)-\log \textsfM_1\right)-\log e \cdot {n_2}g_2\textsfP_1}{\log e\cdot\sqrt{4g_2||x_1^{n_2}(m)||^2+2{n_2}g_2^2\textsfP_1^2}}\label{EQ_r_intermediate20}\\
:=&r_{m,1}(n_2),\label{EQ_def_r1}
\end{align}
where \eqref{EQ_r_intermediate2_0} is from \eqref{EQ_Def_r}.

We further lower bound $r_{m,1}(n_2)$ by substituting the following upper bound:
\begin{align}\label{EQ_power_UB_diff_length}
||x_1^{n_2}(m)||^2\leq ||x_1^{n_1}(m)||^2\leq n_1\textsfP_1,
\end{align}
into the denominator of \eqref{EQ_r_intermediate20}, while we need to ensure the numerator of \eqref{EQ_r_intermediate20} is positive, i.e.,
\begin{align}\label{EQ_r_LB_constraint}
n_2\geq \frac{\log\,\textsfM_1}{\textsfC(g_2\textsfP_1)-\log e\cdot \frac{g_2\textsfP_1}{2(1+g_2\textsfP_1)}}.
\end{align}
We will check the validity of the additional condition \eqref{EQ_r_LB_constraint} at the end of the proof by comparing it to our derived lower bound on $n_2$. Then we can lower bound $r_{m,1}(n_2)$ as follows
\begin{align}
{r}_{m,1}(n_2)
\geq &\frac{[2(1+g_2\textsfP_1)\textsfC(g_2\textsfP_1)-\log e \cdot g_2\textsfP_1]n_2-2(1+g_2\textsfP_1)\log \textsfM_1}{\log e \sqrt{4g_2n_1\textsfP_1+2n_2(g_2)^2\textsfP_1^2}}\notag\\
\geq &\frac{[2(1+g_2\textsfP_1)\textsfC(g_2\textsfP_1)-\log e \cdot g_2\textsfP_1]n_2-2(1+g_2\textsfP_1)\log \textsfM_1}{\log e \sqrt{4g_2\textsfP_1+2(g_2)^2\textsfP_1^2}\sqrt{n_1}}\label{EQ_r_tilde_r0}
\end{align}
for $m\in [1,\, \textsfM_1]$, where \eqref{EQ_r_tilde_r0} is due to $n_1\geq n_2$.

After substituting \eqref{EQ_confusion_prob_2}, \eqref{EQ_UB_given_x}, \eqref{EQ_def_r1}, and \eqref{EQ_r_tilde_r0} into \eqref{EQ_Pe_max_power}, we can derive the following result
\begin{align}
&\frac{1}{\textsf{M}_1}\sum_{m_1=1}^{\textsf{M}_1}\mbox{Pr}(\hat{m}_1\neq m_1\mbox{ at user }2|\,\,m_1 \mbox{ is sent, } \mathcal{C} \mbox{ is used, and use } n_2 \mbox{ symbols to decode})\notag\\
\leq & Q\!\left(\!\!
\frac{[2(1\!+\!g_2\textsfP_1)\textsfC(g_2\textsfP_1)\!-\!\log e \!\cdot g_2\textsfP_1]n_2\!-\!2(1\!+\!g_2\textsfP_1)\log \textsfM_1}{\log e \sqrt{4g_2\textsfP_1+2(g_2)^2\textsfP_1^2}\sqrt{n_1}}
\!\!\right)+c_2\label{EQ_total_Pe2_4}\\
\leq &\epsilon_{SIC1},\label{EQ_total_Pe2_5}
\end{align}
where
\begin{align}
c_2:=&B_0(n_2)+B_1(n_2)
+e^{-\frac{n_2}{2}}+e^{\frac{-n_1^{2a}}{2\textsfP_1}+\ln (2n_1)}\notag\\
=& \frac{2}{\sqrt{n_2}}\left(\frac{\ln 2}{\sqrt{\pi d_1}}+\frac{3}{2}(c_0+c_1\cdot n_2^a)\right) +e^{-\frac{n_2}{2}}+e^{\frac{-n_1^{2a}}{2\textsfP_1}+\ln (2n_1)}\label{EQ_c2_1}\\
\leq &\frac{1}{\sqrt{n_2d_1}} \left(\frac{2\ln 2}{\sqrt{\pi}}+64\sqrt{2}\left((\sqrt{g_2}\textsfP_1)^3+2\textsfP_1
\cdot n_2^a\right)\right) +e^{-\frac{n_2}{2}}+e^{\frac{-n_2^{2a}}{2\textsfP_1}+\ln (2n_2)},
\end{align}
where  \eqref{EQ_c2_1} is from \eqref{EQ_B_2} and \eqref{EQ_Q_Yn_xn}, $d_1:=\left(\frac{\sqrt{g_2}}{1+g_2\textsfP_1}\right)^{3/2}$ and $g_2:=\frac{h_2}{1+h_2\textsfP_2}$. Note that in \eqref{EQ_total_Pe2_5} we enforce the upper bound of the average error probability to be no larger than the target value $\epsilon_{SIC1}$.

Now we further rearrange \eqref{EQ_total_Pe2_4} and \eqref{EQ_total_Pe2_5} by taking the inverse function of $Q$-function as follows
\begin{align}
\frac{[2(1+g_2\textsfP_1)\textsfC(g_2\textsfP_1)-\log e \cdot g_2\textsfP_1]n_2-2(1+g_2\textsfP_1)\log \textsfM_1}{\log e \sqrt{4g_2\textsfP_1+2(g_2)^2\textsfP_1^2}\sqrt{n_1}}
\geq & Q^{-1}\left({\epsilon_{SIC1}} -c_2\right)\label{EQ_r_prime1}\\
=& Q^{-1}\left({\epsilon_{SIC1}} \right)+\mathcal{O}(c_2),\label{EQ_r_prime2}
\end{align}
where \eqref{EQ_r_prime1} is due to the fact that Q-function is monotonically decreasing and \eqref{EQ_r_prime2} is due to the Q-function being continuous so we can apply the Taylor expansion as \cite[(267)]{Polyanskiy_finite_block_length}. By simple algebra, we can solve a lower bound of $n_2$ shown as \eqref{EQ_SIC_constraintED_ED_rate_improvement}. Now compare \eqref{EQ_SIC_constraintED_ED_rate_improvement} and \eqref{EQ_r_LB_constraint}, we find that \eqref{EQ_SIC_constraintED_ED_rate_improvement} is stricter. With a power backoff $\delta$, we complete the proof. \QEDA

\renewcommand{\thesection}{Appendix II}
\section{Proof of the rate region in Theorem \ref{Th_GBC_ED}}\label{APP_Pf_rate_region}

To analyze the weaker user's rate, we modify \eqref{EQ_Pe_max_power} by considering the channel gain $g^{n_1}$ as part of the channel output and in contrast to the analysis in \ref{APP_ED_Gaussian_max_power}, here we use the random coding argument as follows
\begin{align}
\epsilon\leq &P_{ {X}_1^{n_1}}P_{ \tilde{Y}^{n_1}G^{n_1}|{X}_1^{n_1}}\left(i( {X}_1^{n_1}; \tilde{Y}^{n_1})\leq \log\textsfM\right)+\textsf{M}\cdot
\left[P_{ {X}_1^{n_1}}P_{ \tilde{Y}^{n_1}}\left(i( {X}_1^{n_1}; \tilde{Y}^{n_1})\geq \log\textsfM\right)\right]+P_{X_1^{n_1}}\left(X_1^{n_1}\notin \mathcal{F}_{IPC}^{n_1}\right),\label{EQ_Pe_HNOMA}
\end{align}
where
\begin{align}\label{EQ_CAOD_weaker_user}
P_{\tilde{Y}^{n_1}}=\Pi_{j=1}^{n_1} P_{\tilde{Y}_j}\sim\mathcal{N}(\mathbf{0},{\bm\Sigma}),
\end{align}
and
\begin{align}\label{EQ_cov_mat}
  \bm\Sigma:=\left(\begin{array}{cc}
              \bm I_{n_2}\cdot[1+h_1(\textsfP_1+\textsfP_2)] & \bm 0\\
              \bm 0 & \bm I_{n_1-n_2}\cdot[1+h_1\textsfP_1]\\
             \end{array}\right).
\end{align}

Based on \eqref{EQ_CAOD_weaker_user}, we have the following modified information density:
\begin{align}\label{}
  &i(X_1^{n_1};Y^{n_1})\notag\\
  =&\log\left(\frac{(2\pi)^{\frac{-n_1}{2}}(1+h_1\textsfP_2)^{\frac{-n_2}{2}}1^{\frac{-(n_1-n_2)}{2}}\exp\left(\frac{-\sum_{i=1}^{n_2}(Y_{1,i}-\sqrt{h_1}X_{1,i})^2}{2(1+h_1\textsfP_2)}\right)\exp\left(\frac{-\sum_{i=n_2+1}^{n_1}(Y_{1,i}-\sqrt{h_1}X_{1,i})^2}{2}\right)}{(2\pi)^{\frac{-n_1}{2}}(1+h_1(\textsfP_1+\textsfP_2))^{\frac{-n_2}{2}}(1+h_1\textsfP_1)^{\frac{-(n_1-n_2)}{2}}\exp\left(\frac{-\sum_{i=1}^{n_2}Y_{1,i}^2}{2(1+h_1(\textsfP_1+\textsfP_2))}\right)\exp\left(\frac{-\sum_{i=n_2+1}^{n_1}Y_{1,i}^2}{2(1+h_1\textsfP_1)}\right)}\right)\\
  =&n_1\bar{\textsfC}_1+\frac{\log e}{2(1+h_1(\textsfP_1+\textsfP_2))}\left[{h_1}\sum_{i=1}^{n_2}\left(X_{1,i}^2-\frac{\textsfP_1}{1+h_1\textsfP_2}\tilde{Z}_{1,i}^2\right)+2\langle\sqrt{h_1}X_1^{n_2},\tilde{Z}_1^{n_2}\rangle\right]+\notag\\
    & \hspace{5cm}\frac{\log e}{2(1+h_1\textsfP_1)}\left[{h_1}\sum_{i=n_2+1}^{n_1}(X_{1,i}^2-\textsfP_1\tilde{Z}_{1,i}^2)+2\langle\sqrt{h_1}X_{1,n_2+1}^{n_1},\tilde{Z}_{1,n_2+1}^{n_1}\rangle\right],\label{EQ_weaker_info_density3}
  \end{align}
where in \eqref{EQ_weaker_info_density3} we define $\bar{\textsfC}_1$ and ${g}_1$ as in \eqref{EQ_def_bar_C1} and \eqref{EQ_def_g1}, respectively. Based on \eqref{EQ_weaker_info_density3} we can derive the variance of the modified information density as follows. For $i\in[1,\,n_2]$, from \eqref{EQ_weaker_info_density3} we can see that
\begin{align}\label{EQ_weaker_user_mean_info_density_1st_step}
  \mathds{E}[i(X_{1,i};Y_i)]=\textsfC\left(\frac{h_1\textsfP_1}{1+h_1\textsfP_2}\right),
\end{align}
and
\begin{align}
  \Var(i(X_{1,i};Y_i))=&\frac{\log^2 e}{4(1+h_1(\textsfP_1+\textsfP_2))^2}\left\{\mathds{E}\left[\left({h_1}\left(X_{1,i}^2-\frac{\textsfP_1}{1+h_1\textsfP_2}\tilde{Z}_{1,i}^2\right)+2\sqrt{h_1}X_{1,i},\tilde{Z}_{1,i}\right)^2\right]\right\}\label{EQ_weaker_user_var_info_density1}\\
  =&\frac{\log^2 e\cdot h_1\textsfP_1}{1+h_1(\textsfP_1+\textsfP_2)},\label{EQ_weaker_user_var_info_density4}
\end{align}
where in \eqref{EQ_weaker_user_var_info_density4} we use the fact that $\mathds{E}[X_{1,i}^2]=\textsfP_1$, $\mathds{E}[X_{1,i}^3]=0$, $\mathds{E}[X_{1,i}^4]=3\textsfP_1^2$, $\mathds{E}[\tilde{Z}_{1,i}^2]=1+h_1\textsfP_2$, $\mathds{E}[\tilde{Z}_{1,i}^3]=0$, and $\mathds{E}[\tilde{Z}_{1,i}^4]=3(1+h_1\textsfP_2)^2$.
On the other hand, if $i\in[n_2+1,\,n_1]$, by setting $\textsfP_2=0$ in \eqref{EQ_weaker_user_var_info_density4}, we have
\begin{align}
\mathds{E}[i(X_{1,i};Y_i)]&=\textsfC(h_1\textsfP_1),\\
  \Var(i(X_{1,i};Y_i))&=\frac{\log^2 e\cdot h_1\textsfP_1}{1+h_1\textsfP_1}.\label{EQ_weaker_user_var_info_density5}
\end{align}
Then from \eqref{EQ_weaker_user_var_info_density4} and \eqref{EQ_weaker_user_var_info_density5} we compare the dispersion is as follows
\begin{align}\label{}
  \bar{\textsfV}_1&=\frac{1}{n_1}\Var\left(i(X_1^{n_1};Y^{n_1})\right)\\
  &=\log^2 e\cdot\left(p\frac{h_1\textsfP_1}{1+h_1(\textsfP_1+\textsfP_2)}+(1-p)\frac{h_1\textsfP_1}{1+h_1\textsfP_1}\right).\label{EQ_weaker_user_dispersion}
  \end{align}
To show the convergence of the Berry-Esseen ratio of the weaker user, in addition to \eqref{EQ_weaker_user_dispersion}, we need to derive an upper bound of the absolute centralized third moment of the information density, which is shown as follows. We denote $W_j$ by the information density of the weaker user at the $j$-th symbol. Then from the above we know that
\begin{align}\label{}
  W_j-\mu_{W_j}=\left\{\begin{array}{ll}
                       \frac{\log e}{2(1+h_1(\textsfP_1+\textsfP_2))}\left({h_1}\left(X_{1,i}^2-\frac{\textsfP_1}{1+h_1\textsfP_2}\tilde{Z}_{1,i}^2\right)+2\sqrt{h_1}X_{1,i},\tilde{Z}_{1,i}\right), & 1\leq j\leq n_2 \\
                       \frac{\log e}{2(1+h_1\textsfP_1)}\left({h_1}\left(X_{1,i}^2-\textsfP_1\tilde{Z}_{1,i}^2\right)+2\sqrt{h_1}X_{1,i},\tilde{Z}_{1,i}\right), & n_2+1\leq j\leq n_1,
                       \end{array}\right.
\end{align}
and we can bound the absolute centralized third moment of the information density as follows:
\begin{align}
  &\sum_{j=1}^{n_1}\mathds{E}_{XY}[|W_j-\mu_{W_j}|^3]\leq\\
   &\frac{9}{8}\log^3 e \cdot \left\{\frac{1}{(1+h_1(\textsfP_1+\textsfP_2))^3}\sum_{j=1}^{n_2}\!\!\!\!\left(\!\!\!\!9\left(\mathds{E}\left[\left|h_1X_{1,j}^2\right|^3\right]\right)\!\!+\mathds{E}\left[\left|\frac{\textsfP_1}{1+h_2\textsfP_2}\tilde{Z}_{1,j}^2\right|^3\right]\right.\right.\!\!\!\!+\!\!
  \left.\mathds{E}\left[\left|2\sqrt{h_1}X_{1,j}\tilde{Z}_{1,j}\right|^3\right]\!\!\right)\notag\\
  &\hspace{1cm}+\frac{1}{(1+h_1\textsfP_1)^3}\sum_{j=n_2+1}^{n_1}\left(9\left(\mathds{E}\left[\left|h_1X_{1,j}^2\right|^3\right]+\mathds{E}\left[\left|\textsfP_1\tilde{Z}_{1,j}^2\right|^3\right]\right)+\left.\left.\mathds{E}\left[\left|2\sqrt{h_1}X_{1,j}\tilde{Z}_{1,j}\right|^3\right]\right)\right.\right\}.\label{EQ_weaker_user_abs_centralized_3rd_moment}
  \end{align}
We can easily see that each term of the summation in \eqref{EQ_weaker_user_abs_centralized_3rd_moment} is finite, which can be calculated by the absolute moments of $X_{1,j}^2$ and $\tilde{Z}_{1,j}^2$. Therefore, we can further express \eqref{EQ_weaker_user_abs_centralized_3rd_moment} as follows:
\begin{align}
  \sum_{j=1}^{n_1}\mathds{E}_{XY}[|W_j-\mu_{W_j}|^3]&\leq a\cdot n_2+b(n_1-n_2)\label{EQ_centralized_3rd_moment}\\
  &\leq \left\{\begin{array}{ll}
                 a\cdot n_1, & \mbox{ if } a\geq b \\
                 b\cdot n_1, & \mbox{ else },
               \end{array}\right.
\end{align}
where in \eqref{EQ_centralized_3rd_moment}, we collect the coefficients of $n_2$ and $n_1-n_2$ as $a$ and $b$, respectively, where $a>0$ and $b>0$. Then \eqref{EQ_weaker_user_abs_centralized_3rd_moment} is a linear scale of $n_1$. After dividing \eqref{EQ_weaker_user_abs_centralized_3rd_moment} by $(\Var(X^{n_1};\tilde{Y}^{n_1}))^{\frac{3}{2}}$, which can be easily seen from \eqref{EQ_weaker_user_dispersion}, we can observe that the Berry-Esseen ratio of the weaker user is upper bounded by $\mathcal{O}(n^{\frac{1}{2}})$. Based on the above derived results, we can follow the same steps in \ref{APP_ED_Gaussian_max_power} to complete the proof.
\QEDA

\renewcommand{\thesection}{Appendix III}
\section{Proof of $\Pr\left(\sum_{j=1}^{n_2}\left(X_{1,j}+X_{2,j}\right)^2+\sum_{j=n_2+1}^{n_1}X_{1,j}^2>n_1\textsfP_T\right)\leq e^{-\mathcal{O}(n_2)}$}\label{APP_Pr_cross_term_violation}

To simplify the expression, we define $\tilde{X}_{1,j}:=X_{1,j}/\sqrt{\bar{\textsfP}_{11}}$, $\tilde{X}_{2,j}:=X_{2,j}/\sqrt{\bar{\textsfP}_{2}}$, and $Z_j:=\tilde{X}_{1,j}\tilde{X}_{2,j}$ where $\tilde{X}_{1,j},\,\tilde{X}_{2,j}\in\mathcal{N}(0,1)$, $\tilde{X}_{1,j}\ind \tilde{X}_{2,j}$, $j=1,\cdots, n_2$. Then
\begin{align}\label{}
  {\normalfont\mbox{Pr}}\left(\sum_{j=1}^{n_2}X_{1,j}X_{2,j}\geq n_2\delta\right) = {\normalfont\mbox{Pr}}\left(\sum_{j=1}^{n_2}Z_j\geq n_2\cdot t\right),
\end{align}
where $t:=\frac{\delta}{\sqrt{\bar{\textsfP}_{11}\bar{\textsfP}_{2}}}.$ We can derive the moment generating function of $Z_j$ as follows:
\begin{align}\label{EQ_MG}
  \mathds{E}\left[e^{\lambda Z_j}\right]=\frac{1}{\sqrt{1-\lambda ^2}}.
\end{align}

By Chernoff bound, we can derive the following:
\begin{align}\label{EQ_Chernoff_bound_cross}
   {\normalfont\mbox{Pr}}\left(\sum_{j=1}^{n_2}Z_j\geq n_2\cdot t\right)\leq \frac{(1-\lambda^2)^{\frac{-n_2}{2}}}{e^{\lambda n_2 t}}.
\end{align}
By properly selecting an upper bound of the RHS of \eqref{EQ_Chernoff_bound_cross}, e.g.,
\begin{align}\label{EQ_properUB_subExp}
  \frac{1}{\sqrt{1-\lambda^2}}\leq e^{2\lambda^2},
\end{align}
where $|\lambda|<0.8$, the RHS of \eqref{EQ_properUB_subExp} fulfills the definition of sub-exponential distribution \cite[Def. 2.2]{Wainwright_high_dim} with the parameters $(\nu,b)=(2,\,1/0.8)$. This is because $|\lambda|<\frac{1}{b}$ by definition, which can be proved by simple calculus. Then we can invoke the sub-exponential tail bound \cite[(2.20)]{Wainwright_high_dim} to derive \eqref{EQ_cross_Pout} as follows:
\begin{align}\label{EQ_input_cross_violation}
  {\normalfont\mbox{Pr}}\left(\sum_{j=1}^{n_2}Z_j\geq n_2\cdot t\right)\leq e^{\frac{-n_2\cdot t^2}{2\nu^2}}=e^{\frac{-n_2\cdot\delta^2}{8\bar{\textsfP}_{11}\bar{\textsfP}_2}}.
\end{align}

Now we consider the outage of \eqref{EQ_SPC_original_def} with the definition $A_{11}:=\sum_{j=1}^{n_2}X_{1,j}^2,\,A_{2}:=\sum_{j=1}^{n_2}X_{2,j}^2,\,C':=2\sum_{j=1}^{n_2}X_{1,j}X_{2,j}-2n_2\delta,$ and $A_{12}:=\sum_{j=n_2+1}^{n_1}X_{1,j}^2$, as follows:
\begin{align}\label{}
  &\Pr\left(\sum_{j=1}^{n_2}\left(X_{1,j}+X_{2,j}\right)^2+\sum_{j=n_2+1}^{n_1}X_{1,j}^2>n_1\textsfP_T\right)\notag\\
  &\hspace{0.5cm}:=\Pr\left(A_{11}+A_2+A_{12}+C'>n_1\textsfP_T-2n_2\delta\right)\label{EQ_SPC1}\\
  &\hspace{0.5cm}=  \Pr\left( A_{11}+A_2+A_{12}+C'>n_1\textsfP_T-2n_2\delta\mbox{ and }C'>0\right)+\notag\\
  &\hspace{6cm}\Pr\left(A_{11}+A_2+A_{12}+C'>n_1\textsfP_T-2n_2\delta\mbox{ and } C'\leq 0\right)\label{EQ_SPC2}\\
  &\hspace{0.5cm}\leq  \Pr\left( C'>0\right)+
  \Pr\left(A_{11}+A_2+A_{12}+C'>n_1\textsfP_T-2n_2\delta\mbox{ and } C'\leq 0\right)\label{EQ_SPC2_2}\\
  &\hspace{0.5cm}=  \Pr\left( C'>0\right)+\int_{c'\in\mbox{supp}(C')\mbox{ and }c'\leq 0}\Pr\left(A_{11}+A_2+A_{12}>n_1\textsfP_T-2n_2\delta+|c'| \right)d F_{C'}(c')\label{EQ_SPC3}\\
  &\hspace{0.5cm}\leq e^{\frac{-n_2\cdot\delta^2}{8\bar{\textsfP}_{11}}}+\int_{c'\in\mbox{supp}(C')\mbox{ and }c'\leq 0}\Pr\left(A_{11}+A_2+A_{12}>n_1\textsfP_T-2n_2\delta+|c'| \right)d F_{C'}(c'),\label{EQ_SPC4}
  \end{align}
where in \eqref{EQ_SPC1} we define $C':=C-2n_2\delta$, in \eqref{EQ_SPC2} less conditions lead to larger probabilities, in \eqref{EQ_SPC4} we use \eqref{EQ_input_cross_violation} to upper bound the first term in \eqref{EQ_SPC3}. Now define $A_{11}':=A_{11}-n_2\textsfP_{11}+\frac{2n_2\delta}{3}$, $A_{12}':=A_{12}-(n_1-n_2)\textsfP_{12}+\frac{2n_2\delta}{3}$, $A_{2}':=A_{2}-n_2\textsfP_{2}+\frac{2n_2\delta}{3}$. Then we can further express the integrand in \eqref{EQ_SPC3} as follows:
\begin{align}
  \Pr\left(A_{11}+A_2+A_{12}>n_1\textsfP_T-2n_2\delta + |c'| \right)
  &\leq\Pr(A_{11}'+A_2'+A_{12}'>|c'| )\label{EQ_SPC3_3}\\
  &\leq\Pr(A_{11}'+A_2'+A_{12}'>0 )\label{EQ_SPC3_4_2}\\
  &\leq \Pr(A_{11}'>0)+\Pr(A_2'>0)+\Pr(A_{12}'>0),\label{EQ_SPC5}
\end{align}
where \eqref{EQ_SPC3_3} is due to the definitions of $A_{11},\,A_{12},\,A_2$ with the inequality \eqref{EQ_SPC_new}, in \eqref{EQ_SPC5} we recursively use \eqref{EQ_SPC2} and \eqref{EQ_SPC2_2}. Combine \eqref{EQ_SPC4} and \eqref{EQ_SPC5} with the fact that with a proper power backoff, \eqref{EQ_SPC5} can be upper bounded by $e^{-\mathcal{O}(n_2)}$ from the concentration inequality \cite[(4.3)]{Laurent_ChiSquare_Tail}, we have \eqref{EQ_cross_Pout}. To derive \eqref{EQ_cross_Pout2}, we can follow the same steps to derive \eqref{EQ_SPC5} with a slight modification, which completes the proof.
\QEDA.

\renewcommand{\thesection}{Appendix IV}
\section{Proof of Proposition \ref{Lemma_HNOMA_sum_power}}\label{APP_HNOMA_Sum_Power}

We first rearrange \eqref{EQ_SPC_new} as follows
\begin{align}\label{EQ_sum_power_condition}
  p\textsfP_{1,1}+(1-p)\textsfP_{1,2}\leq \textsfP_T-p\textsfP_2.
\end{align}
We then generalize the dispersion expression with IPC in \eqref{EQ_weaker_user_dispersion} to the following form:
\begin{align}\label{EQ_dispersion_sum_power}
  \bar{\textsfV}_1'= \log^2 e\cdot \left(p\frac{g_1'\textsfP_{1,1}}{1+g_1'\textsfP_{1,1}}+(1-p)\frac{g_2'\textsfP_{1,2}}{1+g_2'\textsfP_{1,2}}\right),
\end{align}
where $g_1':=\frac{h_1}{1+h_1\textsfP_2}$ and $g_2':=h_1$ are the equivalent channel gains when the indices of code symbols are from 1 to $n_2$ and from $n_2+1$ to $n_1$, respectively. With a power backoff $\delta$, we can derive $\textsfbarV_1'$ as \eqref{EQ_def_bar_V1_p}. Similarly, we can derive $\textsfbarC'$ from \eqref{EQ_weaker_user_mean_info_density_1st_step} as \eqref{EQ_def_bar_C1_p}, which completes the proof. \QEDA
\bibliographystyle{IEEEtran}
\renewcommand{\baselinestretch}{2}
\bibliography{IEEEabrv,SecrecyPs2_2,../SecrecyPs2,../SecrecyPs22}

\begin{thebibliography}{10}
\providecommand{\url}[1]{#1}
\csname url@samestyle\endcsname
\providecommand{\newblock}{\relax}
\providecommand{\bibinfo}[2]{#2}
\providecommand{\BIBentrySTDinterwordspacing}{\spaceskip=0pt\relax}
\providecommand{\BIBentryALTinterwordstretchfactor}{4}
\providecommand{\BIBentryALTinterwordspacing}{\spaceskip=\fontdimen2\font plus
\BIBentryALTinterwordstretchfactor\fontdimen3\font minus
  \fontdimen4\font\relax}
\providecommand{\BIBforeignlanguage}[2]{{%
\expandafter\ifx\csname l@#1\endcsname\relax
\typeout{** WARNING: IEEEtran.bst: No hyphenation pattern has been}%
\typeout{** loaded for the language `#1'. Using the pattern for}%
\typeout{** the default language instead.}%
\else
\language=\csname l@#1\endcsname
\fi
#2}}
\providecommand{\BIBdecl}{\relax}
\BIBdecl

\bibitem{PHL_ISIT21}
P.-H. Lin, S.-C. Lin, and E.~A. Jorswieck, ``Early decoding for {G}aussian
  broadcast channels with heterogeneous blocklength constraints,'' in
  \emph{Proc. IEEE Int. Symp. Inf. Theory (ISIT) 2021, Melbourne, Australia},
  2021.

\bibitem{PHL_ICC22_ED}
P.-H. Lin, S.-C. Lin, P.-W. Chen, M.~Mross, and E.~A. Jorswieck, ``Rate region
  of {G}aussian broadcast channels with heterogeneous blocklength
  constraints,'' in \emph{Proc. IEEE Int. Conf. on Communi. (ICC) 2022, Seoul,
  Korea}, 2022.

\bibitem{3GPP_URLLC}
3GPP, ``Study on scenarios and requirements for next generation access
  technologies,'' \emph{Technical Report 38.913, Release 14}, Oct. 2016.

\bibitem{3GPP_URLLC2}
------, ``Summary of email discussion on the link level evaluation for {LTE
  URLLC},'' \emph{Technical Report, TSG RAN WG1 Meeting No.92, R1---1801385},
  Mar. 2018.

\bibitem{Durisi_URLLC}
G.~{Durisi}, T.~{Koch}, and P.~{Popovski}, ``Toward massive, ultrareliable, and
  low-latency wireless communication with short packets,'' \emph{Proceedings of
  the IEEE}, vol. 104, no.~9, pp. 1711--1726, 2016.

\bibitem{Polyanskiy_finite_block_length}
Y.~Polyanskiy, H.~V. Poor, and S.~Verd\'{u}, ``Channel coding rate in the
  finite blocklength regime,'' \emph{{IEEE} Trans. Inf. Theory}, vol.~56,
  no.~5, pp. 2307--2359, May 2010.

\bibitem{Ebra_IT15}
E.~{MolavianJazi} and J.~N. {Laneman}, ``A second-order achievable rate region
  for {G}aussian multi-access channels via a central limit theorem for
  functions,'' \emph{{IEEE} Trans. Inf. Theory}, vol.~61, no.~12, pp.
  6719--6733, 2015.

\bibitem{Scarlett_GMAC_DMS}
J.~Scarlett and V.~Y.~F. Tan, ``Second-order asymptotics for the {G}aussian
  {MAC} with degraded message sets,'' \emph{{IEEE} Trans. Inf. Theory},
  vol.~61, no.~12, pp. 6700--6718, 2015.

\bibitem{Tan_MUC_dispersion}
V.~Y.~F. {Tan} and O.~{Kosut}, ``On the dispersions of three network
  information theory problems,'' \emph{{IEEE} Trans. Inf. Theory}, vol.~60,
  no.~2, pp. 881--903, 2014.

\bibitem{Gorce_GBC_dispersion}
A.~{\"{U}nsal} and J.~{Gorce}, ``The dispersion of superposition coding for
  {G}aussian broadcast channels,'' in \emph{Proc. IEEE Information Theory
  Workshop (ITW)}, 2017, pp. 414--418.

\bibitem{Tan_strong_converse_GBC}
S.~L. Fong and V.~Y.~F. Tan, ``A proof of the strong converse theorem for
  {G}aussian broadcast channels via the {G}aussian {P}oincar\'{e} inequality,''
  \emph{{IEEE} Trans. Inf. Theory}, vol.~63, no.~12, pp. 7737--7746, 2017.

\bibitem{Sheldon_GBC_FBL}
P.~Sheldon, D.~Tuninetti, and B.~Smida, ``The {G}aussian broadcast channels
  with a hard deadline and a global reliability constraint,'' in \emph{Proc.
  IEEE Int. Conf. on Communi. (ICC)}, 2021, pp. 1--6.

\bibitem{Scarlett_GPC}
J.~Scarlett, ``On the dispersions of the {G}el'fand-{P}insker channel and dirty
  paper coding,'' \emph{{IEEE} Trans. Inf. Theory}, vol.~61, no.~9, pp.
  4569--4586, Sept. 2015.

\bibitem{Korrai_5G}
P.~K. Korrai, E.~Lagunas, A.~Bandi, S.~K. Sharma, and S.~Chatzinotas, ``Joint
  power and resource block allocation for mixed-numerology-based 5{G} downlink
  under imperfect {CSI},'' \emph{IEEE Open Journal of the Communications
  Society}, vol.~1, pp. 1583--1601, 2020.

\bibitem{xu_hybrid_NOMA}
Y.~{Xu}, C.~{Shen}, T.~{Chang}, S.~{Lin}, Y.~{Zhao}, and G.~{Zhu},
  ``Transmission energy minimization for heterogeneous low-latency {NOMA}
  downlink,'' \emph{{IEEE} Trans. Wireless Commun.}, vol.~19, no.~2, pp.
  1054--1069, 2020.

\bibitem{Tuninetti_GBC_hard_deadline}
D.~{Tuninetti}, B.~{Smida}, N.~{Devroye}, and H.~{Seferoglu}, ``Scheduling on
  the {G}aussian broadcast channel with hard deadlines,'' in \emph{Proc. IEEE
  Int. Conf. on Communi. (ICC)}, 2018, pp. 1--7.

\bibitem{Azarian_ED}
K.~Azarian, H.~E. Gamal, and P.~Schniter, ``On the achievable
  diversity-multiplexing tradeoff in half-duplex cooperative channels,''
  \emph{{IEEE} Trans. Inf. Theory}, vol.~51, no.~12, pp. 4152--4272, Dec. 2005.

\bibitem{Jovicic_CR}
A.~Jovicic and P.~Viswanath, ``Cognitive radio: an information-theoretic
  perspective,'' \emph{{IEEE} Trans. Inf. Theory}, vol.~55, no.~9, pp.
  3945--3958, Sep. 2009.

\bibitem{Hou_EarlyDecoding}
J.~{Hou} and G.~{Kramer}, ``Short message noisy network coding with a
  decode-forward option,'' \emph{{IEEE} Trans. Inf. Theory}, vol.~62, no.~1,
  pp. 89--107, 2016.

\bibitem{Sahin_ED}
C.~{Sahin}, L.~{Liu}, and E.~{Perrins}, ``Early decoding for transmission over
  finite transport blocks,'' in \emph{Proc. IEEE International Symposium on
  Information Theory (ISIT)}, 2014, pp. 1558--1562.

\bibitem{PHL_arxiv21_J_ED}
P.-H. Lin, S.-C. Lin, P.-W. Chen, M.~Mross, and E.~A. Jorswieck, ``{G}aussian
  broadcast channels in heterogeneous blocklength constrained networks,''
  \emph{arxiv:2109.07767}, 2021.

\bibitem{Wainwright_high_dim}
M.~Wainwright, \emph{High-Dimensional Statistics: A Non-Asymptotic Viewpoint},
  1st~ed.\hskip 1em plus 0.5em minus 0.4em\relax Cambridge university press,
  2019.

\bibitem{Laurent_ChiSquare_Tail}
\BIBentryALTinterwordspacing
B.~Laurent and P.~Massart, ``{Adaptive estimation of a quadratic functional by
  model selection},'' \emph{The Annals of Statistics}, vol.~28, no.~5, pp. 1302
  -- 1338, 2000. [Online]. Available:
  \url{https://doi.org/10.1214/aos/1015957395}
\BIBentrySTDinterwordspacing

\end{thebibliography}

\end{document}